\documentclass[10pt]{article}

\PassOptionsToPackage{square,sort,comma,numbers}{natbib}

\usepackage[final]{neurips_2024}

\usepackage{times}
\usepackage{latexsym}
\usepackage{booktabs}
\usepackage{comment}
\usepackage{bbm}
\usepackage{amsmath}
\usepackage{color-edits}
\addauthor{vc}{blue}
\addauthor{ub}{red}
\addauthor{kc}{purple}
\addauthor{ms}{brown}
\usepackage[linewidth=1pt]{mdframed}
\usepackage[utf8]{inputenc}

\usepackage{microtype}

\usepackage{inconsolata}

\usepackage{graphicx}

\usepackage[utf8]{inputenc} % allow utf-8 input
\usepackage[T1]{fontenc}    % use 8-bit T1 fonts
\usepackage{hyperref}       % hyperlinks
\usepackage{url}            % simple URL typesetting
\usepackage{booktabs}       % professional-quality tables
\usepackage{amsfonts}       % blackboard math symbols
\usepackage{nicefrac}       % compact symbols for 1/2, etc.
\usepackage{microtype}      % microtypography
\usepackage{xcolor}         % colors

\title{Modulating Language Model Experiences through Frictions}

\author{
 \textbf{Katherine M. Collins\textsuperscript{1}},
 \textbf{Valerie Chen\textsuperscript{2}},
 \textbf{Ilia Sucholutsky\textsuperscript{3}},
 \textbf{Hannah Rose Kirk\textsuperscript{4}},
\\
 \textbf{Malak Sadek\textsuperscript{5}},
 \textbf{Holli Sargeant\textsuperscript{1}},
 \textbf{Ameet Talwalkar\textsuperscript{2}},
 \textbf{Adrian Weller \textsuperscript{1,6}},
 \textbf{Umang Bhatt\textsuperscript{6,7}}
\\
\\
 \textsuperscript{1}University of Cambridge,
 \textsuperscript{2}Carnegie Mellon University,
 \textsuperscript{3}Princeton University,
 \textsuperscript{4}University of Oxford,
 \\
 \textsuperscript{5}Imperial College,
 \textsuperscript{6}The Alan Turing Institute,
 \textsuperscript{7}New York University
\\
   \textbf{Correspondence:} \href{mailto:kmc61@cam.ac.uk}{kmc61@cam.ac.uk} \href{mailto:umangbhatt@nyu.edu}{umangbhatt@nyu.edu} 
 }

\begin{document}
\maketitle
\begin{abstract}

Language models are transforming the ways that their users engage with the world. Despite impressive capabilities, over-consumption of language model outputs risks propagating unchecked errors in the short-term and damaging human capabilities for critical thinking in the long-term. How can we develop scaffolding around language models to curate more appropriate use? 
We propose \textit{selective frictions} for language model experiences, inspired by behavioral science interventions, to dampen misuse. 
Frictions involve small modifications to a user's experience, e.g., the addition of a button impeding model access and reminding a user of their expertise relative to the model. 
Through a user study with real humans, we observe shifts in user behavior from the imposition of a friction over LLMs in the context of a multi-topic question-answering task as a representative task that people may use LLMs for, e.g., in education and information retrieval. 
We find that frictions modulate over-reliance by driving down users' click rates while minimally affecting accuracy for those topics. Yet, frictions may have unintended effects. We find marked differences in users' click behaviors even on topics where frictions were not provisioned. Our contributions motivate further study of human-AI behavioral interaction to inform more effective and appropriate LLM use.

\end{abstract}

% \section{Outline}
% \begin{enumerate}
%     \item Introduction to frictions in AI system interaction and relevance to LLMs
%     \item Related Work: Cognitive Forcing Functions, Microboundaries, Affordances
%     \item Types/Examples of Frictions wrt Tech/AI: Safety [Graphic Content or Toxic Language], Cost [Indiv or Org], Reinforcement [Indiv you are good at this], or Abstention [Illegal or Uncertainty for Org]
%     \item Case Studies for LLMs: Examples of LLM frictioned interactions [Two column -- pre and post]
%     \item Call to Action: Experiments for the Community to Run
% \end{enumerate}

% \section{Outline}
% \begin{enumerate}
%     \item Introduction to frictions in AI system interaction and relevance to LLMs
%     \item Related Work: Cognitive Forcing Functions, Microboundaries, Affordances
%     \item Types/Examples of Frictions wrt Tech/AI: Safety [Graphic Content or Toxic Language], Cost [Indiv or Org], Reinforcement [Indiv you are good at this], or Abstention [Illegal or Uncertainty for Org] -- \textit{ground specifically in LLMs --- and emphasize efficiency point wrt number of queries (see EMNLP CFP)}
%     \item Case Studies of LLM frictioned interactions --- with real humans -- caveat that the work is preliminary, in one setting, but already quite rich - motivating future work.
%     \item Call to Action: Experiments for the Community to Run; we care not just about highly capable efficient models -- but the \textit{infrastructure around these models' use}
% \end{enumerate}

\section{Introduction}

There is a colloquial adage that ``just because you can, does not mean you should.'' Large language models (LLMs) have seen unprecedented rates of use: OpenAI's ChatGPT had 100 million users within the first two months of release~\citep{raman2023adoption}. However, characterizing regimes of appropriate use is non-trivial: LLMs are general-purpose technologies with a plethora of use cases. LLMs may not be appropriate to deploy in all contexts as we have seen LLMs perform poorly at mathematics and arithmetic~\citep{frieder2023maths, collins2024evaluating}, avoiding biased and hateful statements~\citep{guo2023bias}, and debugging code~\citep{sobania2023debugging}. To better modulate when LLMs are used, we study the \emph{selective} use of LLMs, thereby limiting access to their responses for specific queries, for specific users.

Mechanisms like reinforcement learning with human feedback~\cite{ouyang2022training} and direct preference optimization~\cite{rafailov2024direct} take steps to steer LLM responses away from illegal, undesired, and toxic content. However, there are reasons beyond safety why it could be desirable to curb LLM use. 
In some contexts, discouraging the use of LLMs for particular users can yield economic or personal benefits. 
For example, encouraging students to solve problems on their own instead of accessing answers provided by LLMs may encourage a deeper understanding of and engagement with educational material~\citep{zhu1987learning, schank2013learning}.
To prevent over-reliance on LLMs, dubbed ``algorithm appreciation''~\cite{logg2019algorithm}, we advocate for thoughtful interactions with LLMs where users are vigilant about \textit{when} they use these tools~\citep{zerilli2022transparency}.
% We challenge the community to be deliberate on when to caveat or revoke LLM responses.  

\begin{figure}[t]
    \centering
    \includegraphics[width=0.4\linewidth]{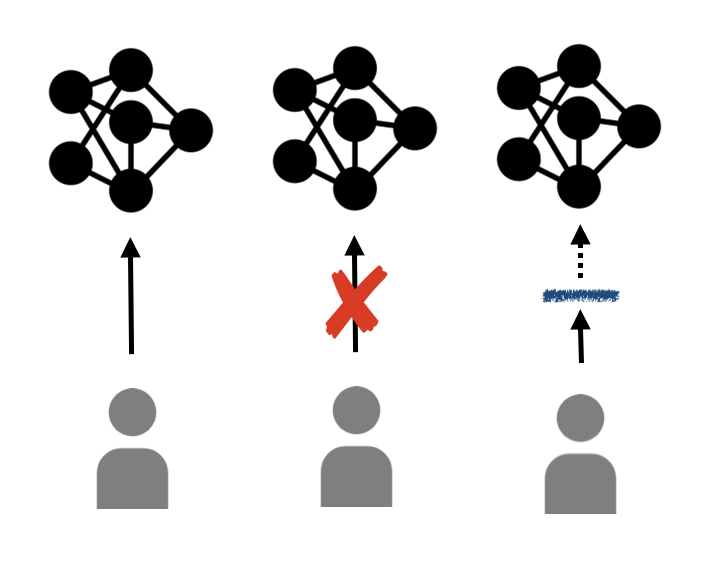}
    \caption{\textbf{Frictions permit continued model access, but require more effort to procure access}. Left: unrestricted access; middle: restricted access; right: frictioned access. 
    We explore the use of selective frictions with respect to user expertise as a way to modulate the ease of model access across task instances.
    % Frictions can be employed generally (to all users for all task instances) or selectively (only for some task instances and/or some specific users).
    }
    \label{fig:access-schematic}
    \vspace{-0.2cm}
\end{figure}

To promote disuse, a spartan option is to restrict access to the LLM response entirely. For example, in deferral schemes, models abstain from providing predictions on specific task instances~\citep{madras2018predict,mozannar2020consistent}. 
% The reasons for deferring on specific tasks, queries, or data may vary. Those interacting with LLMs may have abilities, preferences, and constraints they wish to consider independent of the LLM (e.g., the individual is an expert mathematician). Alternatively, the decision to abstain could depend on model behavior (e.g., the LLM is uncertain about a particular step when proving a theorem), the cost of using certain models (e.g., several advanced LLM models are often hidden behind paywalls), or the broader socio-technical ecosystem surrounding the LLM, whether environmental (e.g. the Carbon costs of running LLM queries~\citep{bannour2021carbon}) or regulatory (e.g. ensuring the use of an LLM does not fall into the `unacceptable risk' category in compliance with the EU AI Act).
As opposed to strict disuse wherein the LLM response is hidden, which could impair user freedoms, we consider selectively adding \textit{friction} to an individual's experience with an LLM. Adapting a definition from~\citet{etzioni2016friction}, we define a \textbf{friction} in the context of LLM assistance as:
\begin{quote}
\small
    \textit{A deliberate design element for increasing the time, effort, or cognitive load of accessing an AI-generated output by prompting conscious consideration of the task at hand.}
\end{quote}

%A friction creates a micro-boundary [cite] that interrupts users' flow. Such a barrier prompts the user to pause momentarily by increasing the resources required to access LLM-generated outputs. The additional resources required could include (a) time taken to see the output, (b) effort (e.g. number of clicks), or (c) cognitive load (e.g. raising awareness on the drawbacks of the output). The introduction of this friction may not prevent users from accessing the LLM-generated output, but would break their flow and encourage conscious and informed engagement with the LLM.

Also referred to as a nudge~\citep{Wilk1999nudges}, microboundary~\citep{cox2016microboundry}, or cognitive forcing~\citep{croskerry2003cognitive},
frictions assuage algorithm appreciation and promote  vigilant use of LLMs: users are encouraged to think twice before relying on an LLM. Similar to selecting when to abstain, introducing selective frictions can depend on model behavior, human expertise, or sociotechnical factors~\cite{bhatt2024should}. For instance, a friction could be selectively applied for a user who is relatively stronger than the LLM at some topic like mathematics.

% For example, frictions can encourage individuals to think twice before relying on an LLM.,
% While friction does not prevent an individual from accessing the LLM output, it can encourage a cognitive dissonance that manifests as a nudge~\citep{Wilk1999nudges} or a microboundary~\citep{cox2016microboundry}. 
% Selectively applying frictions to a user's experience with an LLM would empower a user to only use an LLM for questions in courses where they are struggling: Alice may benefit from LLM access for medieval history but receives frictioned access for geometry, reminding her that she aced her last geometry exam.
% Similar to selecting when to abstain, introducing selective frictions can depend on model behavior, human expertise, or sociotechnical factors~\cite{bhatt2024should}. 
% Such frictions could look like a reminder that the LLM is inaccurate at mathematics queries based on the uncertainty in the LLM's response, or a nudge that the LLMs response contains toxic language. 
% Frictions help codify the colloquial adage ``just because you can, does not mean you should.''
%We demonstrate how frictions can work in practice. We consider two tasks: one closed world, where the response set is pre-specified (i.e., multiple choice), and one open world, where the response is free-form. [to describe...]
%For each task, we run experiments where individuals have full access to the LLM, no access to the LLM, limited access to the LLM, and \textit{frictioned} access to the LLM. We consider three types of friction...

We contribute a case study of the imposition of selective frictions, focusing on selectivity with respect to user expertise. 
% No user is the same; differing in their trust in the model outputs~\citep{buccinca2021trust, zerilli2022transparency}, motivations for use~\citep[i.e.,][]{giglio2023use}, and expertise in prompting and troubleshooting~\citep{}.
% Blanket interventions for all users may preferentially favor some subsets of users. As such, frictions should be, to an extent, tailored to the user and context. 
Specifically, we consider a question-answer setting, reminiscent of ``information-seeking'', knowledge retrieval tasks that users may engage in with AI-based search summaries, e.g., Perplexity or Google Summaries. 
% Specifically, we consider a subset of questions from the popular and challenging NLP benchmark, MMLU~\citep{hendryckstest2021}. 
We extend the user interface, \texttt{Modiste}, from \citet{bhatt2023learning} to explore the imposition of an extra-click \textit{selectively} (on only some topics, for some users) before the user can receive LLM assistance. 
We explore the impact of friction on users' click rates and attainable accuracy on multiple choice question from the popular and challenging NLP benchmark, MMLU~\citep{hendryckstest2021} and other auxiliary measures, such as a user's confidence in their performance and that of the LLM. 
We observe marked behavior in users' click rates from the introduction of friction, providing initial evidence that frictioning LLM access can serve as one effective ``lever'' to design to modulate user experiences and help titrate overreliance. 
Yet, our study urges caution---to our surprise, we observe potential ``spillover'' effects where user behavior changes (i.e., reduced LLM engagements) even when no friction was imposed. 
Our study motivates further interdisciplinary human-centric work studying the interplay of human behavior, pragmatic inferences, and LLM predictions.

\section{A Case Study in Selective Frictions}

% We begin to explore the design and deployment of selective frictions for LLM experiences. Specifically, we consider assisted question-answering. %[.... link up with search?]

\subsection{Task} 
We begin to explore the design and deployment of selective frictions for LLM experiences. Specifically, we consider assisted question-answering. Prior studies including~\citet{mozannar2023effective,bhatt2023learning} have explored LLM assistance in answering multiple-choice questions from MMLU~\citep{hendryckstest2021}; in particular, we build on the set-up and \texttt{Modiste} interface from~\citet{bhatt2023learning}, which supports rapid prototyping of user studies under various forms of assistance. 
Participants answer a total of 60 multiple-choice questions sampled from four topics of MMLU: US foreign policy, elementary mathematics, high school computer science, and high school biology. 
% We extend the \texttt{Modiste} interface from \citet{bhatt2023learning}, which supports rapid prototyping of user studies under various forms of assistance, to permit the incorporation of frictions.

In the baseline condition, participants can press a button to ``query'' the LLM and receive assistance on the current multiple-choice question (as shown in Figure~\ref{fig:btn-click-interface}), which then highlights one of the four multiple-choice options (as shown in Figure~\ref{fig:btn-click-interface-llm-shown}).
As in many real-world settings, we selected a pool of questions where the model is not always correct, as it may not be in many real-world settings, as discussed in Appendix~\ref{llm-pred-details}. 

\subsection{Instantiation of Selective Friction}

% In \citet{bhatt2023learning}, if users were given access to the model prediction, the model prediction was automatically overlaid on the users' multiple-choice platter. 
% Instead, we conceal the prediction initially, forcing the user to press a button to access the prediction (see Figure~\ref{fig:friction-interface}), similar to the type of friction studied in~\citet{buccinca2021trust}. 
% Such a change enables us to explore when users elect to access the LLM. 
We propose a selective friction on top of this button by presenting the user with a second button requiring them to click again, indicating that they are certain that they want to see the model prediction. 
While the first button can be considered a friction in and of itself~\citep{buccinca2021trust}; rather, we treat it as a baseline to compare \textit{selective frictions}.

\textbf{When to impose a friction?} 
There are a variety of reasons for which a friction could benefit user subgroups, depending on the context. In our case study, we study one characteristic: user expertise. If one is already good at computer science, one may not benefit from access to the LLM prediction, particularly if the model has low accuracy.

To assess user expertise across the MMLU topics, we first have the user take a brief quiz (5 questions for each of the 4 topics). If they achieve higher performance than the LLM in a particular topic, then they will be presented with the friction for all questions of that topic in the ``test'' phase. If the user achieves the same expected topic performance as the model, when they indicate they want to see the model's prediction, there is no friction on access. 
%Details on the LLM performance for each topic are included in Appendix ~\ref{llm-preds}.
We decide to friction a new question $x$ if the following quantity is nonzero:
% [todo -- add equation showing difference in performance]
\begin{equation}
   \texttt{friction}(x) = \mathbbm{1}[\text{LLM}_{t(x)} > \text{User}_{t(x)}]
   \label{eq:fric}
\end{equation}
where $t(x)$ represents the topic of the query at hand, $\text{LLM}_{t(x)}$ represents the expected topic performance of the LLM, and $\text{User}_{t(x)}$ represents the expected topic performance based on the brief 5 question quiz. Details on the LLM performance for each topic are included in Appendix~\ref{llm-pred-details}.

\textbf{How to present the friction?} 
The friction takes the form of clicking a second button to view the LLM prediction. But what should this friction say? Small changes in wording can induce markedly different behavior in humans~\citep{thaler2009nudge,halpern2016inside}. To selectively friction by user expertise, we remind users of their expertise (relative to the model). 
We present performance as a frequency drawing on \citet{lai2019explanations} using the following template: \texttt{Do you really want to see the prediction? The AI model gets an average of X out of 10 questions correct on Math. Based on your warmup answers, we estimate that you get an average of Y out of 10 questions correct.} 
We present an example interface in Figure \ref{fig:friction-interface}.

\subsection{Participants} We recruit $100$ participants from Prolific~\citep{palan2018prolific} in an institutionally-ethics reviewed study; participants are recruited from the US and required to speak English as a first language. Participants are randomly assigned to either the \textit{selective-friction} or \textit{baseline} condition ($N=47$ and $53$). 
% In the latter, participants never have friction. 
In the friction condition, based on Eq.~\ref{eq:fric}, $42$ of $53$ participants received friction for foreign policy, $49$ for mathematics, $12$ for computer science, and none for biology. We include more details in Appendix ~\ref{sec:human-exp-details}. 

%Participants receive the quiz for all conditions. The same questions are presented across both conditions, selected from three batches of $60$ questions, as in~\citep{bhatt2023learning}. The ``test'' phase involves $10$ questions per topic. Participants are provided feedback as to whether they (and the model, if seen) are correct after each test trial. Feedback is not given in the quiz phase. Participants are paid at a base rate of $\$9/hr$ for an expected $30$ minute experiment with an optional bonus up to $\$10/hr$ for correct answers; we apply the bonus to all participants.

\subsection{Metrics}

We focus on three metrics: (i) user accuracy over the questions for a given topic, (ii) click rate for questions in a given topic, which is the proportion of times that the user clicks to see the LLM prediction for $M$ questions within a topic\footnote{In the frictioned setting, since the user technically needs to click twice before accessing the model, we only tally the second click.}, and at the end of the study, (iii) the users' self-reported belief in their performance, as well as the LLM's performance, on each topic. %We include details of our self-belief questionarre in the Appendix. 

% % defined at the topic level
% \begin{equation}
%    \texttt{click-rate}(t) = \sum_{i=1}^{M}\mathbbm{1}[\texttt{friction(x_i)} \texttt{double_click}]  + \mathbbm{1}[NOT \texttt{friction(x_i)} \texttt{single_click}]
% \end{equation}

\section{Results} 

% We analyze participant accuracy and click rates, which refer to instances where, if the user was presented with a second button before being able to see the prediction (a friction), then a click was marked. 

% We next present key insights from our empirical case study. 

\begin{figure*}
    \centering
    \includegraphics[width=0.8\textwidth]{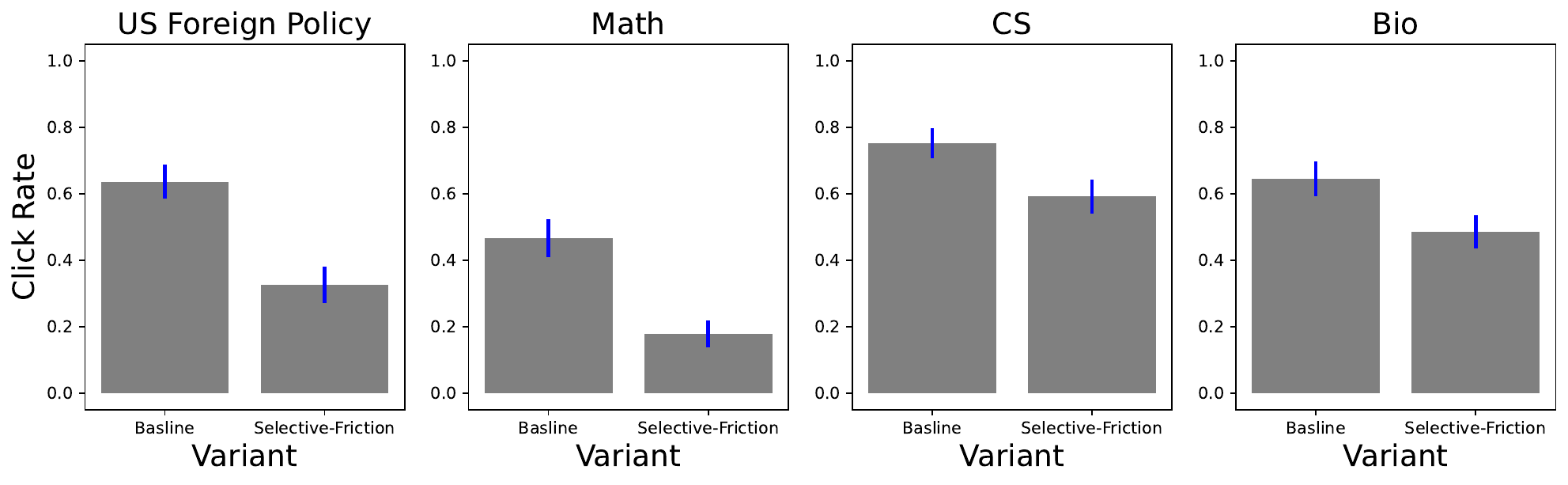}
    \caption{\textbf{Frictioning reduces clicks to see LLM predictions.} We measure the click rate for each user across topics. We find that, for all topics, click rates are statistically significantly reduced ($p < 0.05$) in the selective friction condition. Error bars indicate standard error over participants.}
    \label{fig:num-click-main}
\end{figure*}

\textbf{Key Finding 1: Selective frictions can reduce click rates while maintaining accuracy.} We analyze participant accuracy and click rates and conduct Ordinary Least Squares Regressions with Benjamini-Hochberg correction, using a significance level of $0.05$.
We find that frictioning user experiences, in the way we have done here, indeed significantly lowers user click-through rates as shown in Figure~\ref{fig:num-click-main} ($p<0.05$). 
These results are encouraging, demonstrating friction may be one way to encourage users to solve problems independently. 
This finding is buttressed by minimal, not significant change in the users' accuracy, furthering the benefits of friction to users' critical thinking. % some overreliance

%To our surprise, we do not see a strong change in user accuracy from the imposition of frictions on a users' LLM experiences -- rather, we note that frictions, interestingly, tend to \textit{maintain user accuracy} but from \textit{fewer engagements}. That is, users are able to achieve comparable accuracy despite clicking less. This finding has wide-ranging impliciations -- frictions may be one way to help improve the efficiency of LLM interactions; querying models is costly and can have exorbinant impacts on [...emissions...?]. Going forward we may want to reduce unnecessary queries at inference-time -- we see preliminary evidence that frictions may be one way to do just that, in a targeted fashion.

% \paragraph{Reduce instances of egregious overreliance}

% Further, we see that frictioning can reduce the number of 
% We find that our friction yields a reduction in the absolute number of instances of overreliance that a participants makes. 

\textbf{Key Finding 2: Frictions can induce unintentional spillover effects.} However, to our surprise, we see that click rates drop for participants in the friction condition for biology---\textit{even though no participants were frictioned specifically for biology}. This observation is important; frictioning users' experience in one region of the task space may influence users' decisions in other regions. We speculate why this may be happening, and encourage future work to empirically investigate this phenomenon further. 

When a user is frictioned, we inform them of their own performance and that of the model (which, by definition of seeing the friction, is necessarily lower than the users'). 
The user may overgeneralize the lower model behavior on the frictioned topics to non-frictioned topics. 
We do observe a drop in users' predicted model accuracy in Table~\ref{agg_compare}, across all topics. 
Alternatively, or additionally, frictions may encourage a user's own abilities and increase self-confidence in other questions. We correspondingly observe in Table~\ref{agg_compare} that users' self-confidence tends to increase in the frictioned setting. 
%We find a slight increase in time spent on each biology problem in Table \ref{time_spent} in the frictioned settings, suggesting that users may be more motivated to try on their own for longer. 
%However, [...other possibilities?]. 

% \paragraph{Qualitative analysis: when do frictions fail?}

% We find that, in some instances for some participants, frictions are ineffective at reducing user clicks. In particular, we notice that participants who seem to have the lowest self-confidence in their ability 

\section{Discussion}

As users increasingly access powerful AI systems, buttressed by lightweight natural language interfaces, questions around how system designers can encourage and safeguard appropriate use grows more urgent. 
% Frictions offer one lever for designers to modulate usage while preserving general user freedoms.
Our study demonstrates that small changes to user interfaces in the form of frictions can modulate user behavior, while preserving general user freedoms. 
We show that appropriately designed frictions can reduce user engagement and instances of over-reliance with minimal change in user accuracy.
It is thus possible that frictions can serve as a critical tool in promoting the responsible use of LLMs.
By incorporating barriers that encourage users to engage more critically with AI-generated content, policymakers can help ensure that LLMs are used thoughtfully and selectively, thus preserving and fostering users' impartiality and autonomy~\citep{barletta_rapid_2023,dignum2017,sunstein_ethics_2016}.
% This increased transparency is crucial for building accountability into LLM deployment, as users understand when and why they are engaging with LLMs.

Here, we focused on adding hurdles along a user's path to engaging an LLM. Much work in the behavioral sciences has studied \textit{positive interventions} to encourage particular kinds of behavior by ``nudging''~\citep{thaler2009nudge}. 
Next steps can explore nudges in our MMLU and other settings, as well as alternate mechanisms for instantiating selective frictions drawing on computational models of human behavior~\citep{callaway2023optimal}. 
Further, while our selective frictioning design permits tailored user experiences, e.g., by expertise as we have shown, personalization of LLM experiences can come with risks~\citep{kirk2024benefits}.

While we find that our frictions can dampen excess engagement with an LLM when a user has appropriate expertise, we find that targeted frictions can have ``spillover effects'' wherein users' behavior changes even on topics where frictions were not added nor intended to be added. Our preliminary observations urge caution for designers of interventions around AI systems -- human behavior is complex, and small changes in the realm of interaction may ripple into another. We are excited by future research at the intersection of AI and the behavioral sciences towards more effective ``thought partners''~\citep{collins2024building}.

\section*{Limitations}

Our case study focuses on a single type of friction surrounding selective reminders with respect to user and model expertise; future work is needed to explore whether there are more effective frictions in terms of click rate modulation that may reduce spillover effects.
Additionally, from our current study design, we cannot observe whether the user really needed and/or benefited from the LLM prediction. 
For example, some users clicked simply out of curiosity or to double-check their answers as noted in some post-survey responses (see Appendix \ref{post-resps}), which likely reduced the observed effect.  
As we only obtain the user's final prediction, future work might consider an alternative study design that asks the user for their answer before they see the LLM prediction, which may change the user's decision-making process. Another limitation of this work is our focus on a single dataset, MMLU. 
Since some of the tasks in question are quite challenging. 
We do not see any participants achieve high enough biology performance to be frictioned; many people may feel they need support from the LLM regardless. It is possible, as well, that our quiz does not obtain an adequate appraisal of participant expertise. 
We only evaluate participants on $5$ questions per topic; as such, the expertise profile procured is necessarily an estimate -- and we employ a necessarily reductive binary jurisdiction as to whether or not to apply a friction according to this coarse assesment of expertise. 
Future work is needed to explore alternative expertise elicitation schemes, and to understand whether click rate modulation and possible spillover effects in LLM experiences generalize to other settings and user populations.

\bibliographystyle{abbrvnat}
\bibliography{main}

\begin{thebibliography}{53}
\providecommand{\natexlab}[1]{#1}
\providecommand{\url}[1]{\texttt{#1}}
\expandafter\ifx\csname urlstyle\endcsname\relax
  \providecommand{\doi}[1]{doi: #1}\else
  \providecommand{\doi}{doi: \begingroup \urlstyle{rm}\Url}\fi

\bibitem[Alon-Barkat and Busuioc(2023)]{alon2023human}
S.~Alon-Barkat and M.~Busuioc.
\newblock Human--ai interactions in public sector decision making:“automation bias” and “selective adherence” to algorithmic advice.
\newblock \emph{Journal of Public Administration Research and Theory}, 33\penalty0 (1):\penalty0 153--169, 2023.

\bibitem[Barletta et~al.(2023)Barletta, Caivano, Gigante, and Ragone]{barletta_rapid_2023}
V.~Barletta, D.~Caivano, D.~Gigante, and A.~Ragone.
\newblock A rapid review of responsible {AI} frameworks: How to guide the development of ethical {AI}.
\newblock In \emph{Proceedings of the 27th International Conference on Evaluation and Assessment in Software Engineering}, pages 358--367. {ACM}, 2023.
\newblock \doi{10.1145/3593434.3593478}.
\newblock URL \url{https://doi.org/10.1145/3593434.3593478}.

\bibitem[Bhatt and Sargeant(2024)]{bhatt2024should}
U.~Bhatt and H.~Sargeant.
\newblock When should algorithms resign?
\newblock \emph{arXiv preprint arXiv:2402.18326}, 2024.

\bibitem[Bhatt et~al.(2023)Bhatt, Chen, Collins, Kamalaruban, Kallina, Weller, and Talwalkar]{bhatt2023learning}
U.~Bhatt, V.~Chen, K.~M. Collins, P.~Kamalaruban, E.~Kallina, A.~Weller, and A.~Talwalkar.
\newblock Learning personalized decision support policies.
\newblock \emph{arXiv preprint arXiv:2304.06701}, 2023.

\bibitem[Bu\c{c}inca et~al.(2021)Bu\c{c}inca, Malaya, and Gajos]{buccinca2021}
Z.~Bu\c{c}inca, M.~B. Malaya, and K.~Z. Gajos.
\newblock To trust or to think: Cognitive forcing functions can reduce overreliance on ai in ai-assisted decision-making.
\newblock \emph{Proc. ACM Hum.-Comput. Interact.}, 5\penalty0 (CSCW1), apr 2021.
\newblock \doi{10.1145/3449287}.
\newblock URL \url{https://doi.org/10.1145/3449287}.

\bibitem[Bu{\c{c}}inca et~al.(2021)Bu{\c{c}}inca, Malaya, and Gajos]{buccinca2021trust}
Z.~Bu{\c{c}}inca, M.~B. Malaya, and K.~Z. Gajos.
\newblock To trust or to think: cognitive forcing functions can reduce overreliance on ai in ai-assisted decision-making.
\newblock \emph{Proceedings of the ACM on Human-Computer Interaction}, 5\penalty0 (CSCW1):\penalty0 1--21, 2021.

\bibitem[Bu{\c{c}}inca et~al.(2024)Bu{\c{c}}inca, Swaroop, Paluch, Murphy, and Gajos]{buccinca2024towards}
Z.~Bu{\c{c}}inca, S.~Swaroop, A.~E. Paluch, S.~A. Murphy, and K.~Z. Gajos.
\newblock Towards optimizing human-centric objectives in ai-assisted decision-making with offline reinforcement learning.
\newblock \emph{arXiv preprint arXiv:2403.05911}, 2024.

\bibitem[Callaway et~al.(2023)Callaway, Hardy, and Griffiths]{callaway2023optimal}
F.~Callaway, M.~Hardy, and T.~L. Griffiths.
\newblock Optimal nudging for cognitively bounded agents: A framework for modeling, predicting, and controlling the effects of choice architectures.
\newblock \emph{Psychological Review}, 2023.

\bibitem[Chen et~al.(2023)Chen, Liao, Wortman~Vaughan, and Bansal]{chen2023understanding}
V.~Chen, Q.~V. Liao, J.~Wortman~Vaughan, and G.~Bansal.
\newblock Understanding the role of human intuition on reliance in human-ai decision-making with explanations.
\newblock \emph{Proceedings of the ACM on Human-Computer Interaction}, 7\penalty0 (CSCW2):\penalty0 1--32, 2023.

\bibitem[Collins et~al.(2024{\natexlab{a}})Collins, Jiang, Frieder, Wong, Zilka, Bhatt, Lukasiewicz, Wu, Tenenbaum, Hart, et~al.]{collins2024evaluating}
K.~M. Collins, A.~Q. Jiang, S.~Frieder, L.~Wong, M.~Zilka, U.~Bhatt, T.~Lukasiewicz, Y.~Wu, J.~B. Tenenbaum, W.~Hart, et~al.
\newblock Evaluating language models for mathematics through interactions.
\newblock \emph{Proceedings of the National Academy of Sciences}, 121\penalty0 (24):\penalty0 e2318124121, 2024{\natexlab{a}}.

\bibitem[Collins et~al.(2024{\natexlab{b}})Collins, Sucholutsky, Bhatt, Chandra, Wong, Lee, Zhang, Zhi-Xuan, Ho, Mansinghka, et~al.]{collins2024building}
K.~M. Collins, I.~Sucholutsky, U.~Bhatt, K.~Chandra, L.~Wong, M.~Lee, C.~E. Zhang, T.~Zhi-Xuan, M.~Ho, V.~Mansinghka, et~al.
\newblock Building machines that learn and think with people.
\newblock \emph{Nature Human Behaviour}, 8\penalty0 (10):\penalty0 1851--1863, 2024{\natexlab{b}}.

\bibitem[Cox et~al.(2016)Cox, Gould, Cecchinato, Iacovides, and Renfree]{cox2016microboundry}
A.~L. Cox, S.~J. Gould, M.~E. Cecchinato, I.~Iacovides, and I.~Renfree.
\newblock Design frictions for mindful interactions: The case for microboundaries.
\newblock In \emph{Proceedings of the 2016 CHI Conference Extended Abstracts on Human Factors in Computing Systems}, page 1389–1397, 2016.
\newblock \doi{10.1145/2851581.2892410}.
\newblock URL \url{https://doi.org/10.1145/2851581.2892410}.

\bibitem[Croskerry(2003)]{croskerry2003cognitive}
P.~Croskerry.
\newblock Cognitive forcing strategies in clinical decisionmaking.
\newblock \emph{Annals of emergency medicine}, 41\penalty0 (1):\penalty0 110--120, 2003.

\bibitem[Dignum(2017)]{dignum2017}
V.~Dignum.
\newblock Responsible autonomy.
\newblock In \emph{Proceedings of the 26th International Joint Conference on Artificial Intelligence}, IJCAI'17, page 4698–4704. AAAI Press, 2017.
\newblock ISBN 9780999241103.

\bibitem[Etzioni(2016)]{etzioni2016friction}
A.~Etzioni.
\newblock \emph{A Socio-Economic Perspective on Friction}, chapter~3.
\newblock Routledge, 2016.

\bibitem[Frieder et~al.(2023)Frieder, Pinchetti, Griffiths, Salvatori, Lukasiewicz, Chevalier, and Berner]{frieder2023maths}
S.~Frieder, L.~Pinchetti, R.~Griffiths, T.~Salvatori, P.~Lukasiewicz, T.~Petersen, A.~Chevalier, and J.~Berner.
\newblock Mathematical capabilities of chatgpt.
\newblock \emph{arXiv preprint}, 2023.

\bibitem[Geifman and El-Yaniv(2017)]{geifman2017selective}
Y.~Geifman and R.~El-Yaniv.
\newblock Selective classification for deep neural networks.
\newblock \emph{Advances in neural information processing systems}, 30, 2017.

\bibitem[Green and Chen(2019)]{green2019principles}
B.~Green and Y.~Chen.
\newblock The principles and limits of algorithm-in-the-loop decision making.
\newblock \emph{Proceedings of the ACM on Human-Computer Interaction}, 3\penalty0 (CSCW):\penalty0 1--24, 2019.

\bibitem[Guo et~al.(2023)Guo, Zhang, Wang, Jiang, Nie, Ding, Yue, and Wu]{guo2023bias}
B.~Guo, X.~Zhang, Z.~Wang, M.~Jiang, J.~Nie, Y.~Ding, J.~Yue, and Y.~Wu.
\newblock How close is chatgpt to human experts? comparison corpus, evaluation, and detection.
\newblock \emph{arXiv preprint arXiv:2301.07597}, 2023.

\bibitem[Halpern(2016)]{halpern2016inside}
D.~Halpern.
\newblock \emph{Inside the nudge unit: How small changes can make a big difference}.
\newblock Random House, 2016.

\bibitem[Han(2020)]{han2020human}
L.~Han.
\newblock When the human is in the loop: Cost, effort and behavior.
\newblock In \emph{Proceedings of the 43rd International ACM SIGIR Conference on Research and Development in Information Retrieval}, pages 2480--2480, 2020.

\bibitem[Hendrycks et~al.(2021)Hendrycks, Burns, Basart, Zou, Mazeika, Song, and Steinhardt]{hendryckstest2021}
D.~Hendrycks, C.~Burns, S.~Basart, A.~Zou, M.~Mazeika, D.~Song, and J.~Steinhardt.
\newblock Measuring massive multitask language understanding.
\newblock \emph{Proceedings of the International Conference on Learning Representations (ICLR)}, 2021.

\bibitem[Hummel and Maedche(2019)]{hummel2019}
D.~Hummel and A.~Maedche.
\newblock How effective is nudging? a quantitative review on the effect sizes and limits of empirical nudging studies.
\newblock \emph{Journal of Behavioral and Experimental Economics}, 80:\penalty0 47--58, 2019.
\newblock ISSN 2214-8043.
\newblock \doi{https://doi.org/10.1016/j.socec.2019.03.005}.
\newblock URL \url{https://www.sciencedirect.com/science/article/pii/S2214804318303999}.

\bibitem[Jones et~al.(2020)Jones, Sagawa, Koh, Kumar, and Liang]{jones2020selective}
E.~Jones, S.~Sagawa, P.~W. Koh, A.~Kumar, and P.~Liang.
\newblock Selective classification can magnify disparities across groups.
\newblock In \emph{International Conference on Learning Representations}, 2020.

\bibitem[Joshi et~al.(2023)Joshi, Liu, Ramnath, Chan, Tong, Nie, Wang, Choi, and Ren]{joshi2023machine}
B.~Joshi, Z.~Liu, S.~Ramnath, A.~Chan, Z.~Tong, S.~Nie, Q.~Wang, Y.~Choi, and X.~Ren.
\newblock Are machine rationales (not) useful to humans? measuring and improving human utility of free-text rationales.
\newblock In \emph{Proceedings of the 61st Annual Meeting of the Association for Computational Linguistics (Volume 1: Long Papers)}, pages 7103--7128, 2023.

\bibitem[Kirk et~al.(2024)Kirk, Vidgen, R{\"o}ttger, and Hale]{kirk2024benefits}
H.~R. Kirk, B.~Vidgen, P.~R{\"o}ttger, and S.~A. Hale.
\newblock The benefits, risks and bounds of personalizing the alignment of large language models to individuals.
\newblock \emph{Nature Machine Intelligence}, pages 1--10, 2024.

\bibitem[Lai and Tan(2019)]{lai2019explanations}
V.~Lai and C.~Tan.
\newblock On human predictions with explanations and predictions of machine learning models: A case study on deception detection.
\newblock In \emph{Proceedings of the Conference on Fairness, Accountability, and Transparency}, FAT* '19, page 29–38, New York, NY, USA, 2019. Association for Computing Machinery.
\newblock ISBN 9781450361255.
\newblock \doi{10.1145/3287560.3287590}.
\newblock URL \url{https://doi.org/10.1145/3287560.3287590}.

\bibitem[Lai et~al.(2021)Lai, Chen, Liao, Smith-Renner, and Tan]{lai2021towards}
V.~Lai, C.~Chen, Q.~V. Liao, A.~Smith-Renner, and C.~Tan.
\newblock Towards a science of human-ai decision making: a survey of empirical studies.
\newblock \emph{arXiv preprint arXiv:2112.11471}, 2021.

\bibitem[Lai et~al.(2022)Lai, Carton, Bhatnagar, Liao, Zhang, and Tan]{lai2022human}
V.~Lai, S.~Carton, R.~Bhatnagar, Q.~V. Liao, Y.~Zhang, and C.~Tan.
\newblock Human-{AI} collaboration via conditional delegation: A case study of content moderation.
\newblock In \emph{CHI Conference on Human Factors in Computing Systems}, pages 1--18, 2022.

\bibitem[Li et~al.(2024)Li, Lu, and Yin]{li2024decoding}
Z.~Li, Z.~Lu, and M.~Yin.
\newblock Decoding ai's nudge: A unified framework to predict human behavior in ai-assisted decision making.
\newblock \emph{arXiv e-prints}, pages arXiv--2401, 2024.

\bibitem[Logg et~al.(2019)Logg, Minson, and Moore]{logg2019algorithm}
J.~M. Logg, J.~A. Minson, and D.~A. Moore.
\newblock Algorithm appreciation: People prefer algorithmic to human judgment.
\newblock \emph{Organizational Behavior and Human Decision Processes}, 151:\penalty0 90--103, 2019.

\bibitem[Ma et~al.(2023)Ma, Lei, Wang, Zheng, Shi, Yin, and Ma]{ma2023should}
S.~Ma, Y.~Lei, X.~Wang, C.~Zheng, C.~Shi, M.~Yin, and X.~Ma.
\newblock Who should i trust: Ai or myself? leveraging human and ai correctness likelihood to promote appropriate trust in ai-assisted decision-making.
\newblock In \emph{Proceedings of the 2023 CHI Conference on Human Factors in Computing Systems}, pages 1--19, 2023.

\bibitem[Madras et~al.(2018)Madras, Pitassi, and Zemel]{madras2018predict}
D.~Madras, T.~Pitassi, and R.~Zemel.
\newblock Predict responsibly: Improving fairness and accuracy by learning to defer, 2018.

\bibitem[Mejtoft et~al.(2019)Mejtoft, Hale, and S{\"o}derstr{\"o}m]{mejtoft2019design}
T.~Mejtoft, S.~Hale, and U.~S{\"o}derstr{\"o}m.
\newblock Design friction.
\newblock In \emph{Proceedings of the 31st European Conference on Cognitive Ergonomics}, pages 41--44, 2019.

\bibitem[Mozannar and Sontag(2020)]{mozannar2020consistent}
H.~Mozannar and D.~Sontag.
\newblock Consistent estimators for learning to defer to an expert.
\newblock In \emph{International Conference on Machine Learning}, pages 7076--7087. PMLR, 2020.

\bibitem[Mozannar et~al.(2023)Mozannar, Lee, Wei, Sattigeri, Das, and Sontag]{mozannar2023effective}
H.~Mozannar, J.~J. Lee, D.~Wei, P.~Sattigeri, S.~Das, and D.~Sontag.
\newblock Effective human-ai teams via learned natural language rules and onboarding.
\newblock \emph{arXiv preprint arXiv:2311.01007}, 2023.

\bibitem[Mozannar et~al.(2024)Mozannar, Lee, Wei, Sattigeri, Das, and Sontag]{mozannar2024effective}
H.~Mozannar, J.~Lee, D.~Wei, P.~Sattigeri, S.~Das, and D.~Sontag.
\newblock Effective human-ai teams via learned natural language rules and onboarding.
\newblock \emph{Advances in Neural Information Processing Systems}, 36, 2024.

\bibitem[Ouyang et~al.(2022)Ouyang, Wu, Jiang, Almeida, Wainwright, Mishkin, Zhang, Agarwal, Slama, Ray, et~al.]{ouyang2022training}
L.~Ouyang, J.~Wu, X.~Jiang, D.~Almeida, C.~Wainwright, P.~Mishkin, C.~Zhang, S.~Agarwal, K.~Slama, A.~Ray, et~al.
\newblock Training language models to follow instructions with human feedback.
\newblock \emph{Advances in Neural Information Processing Systems}, 35:\penalty0 27730--27744, 2022.

\bibitem[Palan and Schitter(2018)]{palan2018prolific}
S.~Palan and C.~Schitter.
\newblock Prolific. ac—a subject pool for online experiments.
\newblock \emph{Journal of Behavioral and Experimental Finance}, 17:\penalty0 22--27, 2018.

\bibitem[Rafailov et~al.(2024)Rafailov, Sharma, Mitchell, Manning, Ermon, and Finn]{rafailov2024direct}
R.~Rafailov, A.~Sharma, E.~Mitchell, C.~D. Manning, S.~Ermon, and C.~Finn.
\newblock Direct preference optimization: Your language model is secretly a reward model.
\newblock \emph{Advances in Neural Information Processing Systems}, 36, 2024.

\bibitem[Raman et~al.(2023)Raman, Mandal, Das, and et]{raman2023adoption}
R.~Raman, S.~Mandal, P.~Das, and a.~et.
\newblock University students as early adopters of chatgpt: Innovation diffusion study.
\newblock \emph{Research Square preprint}, 2023.
\newblock \doi{https://doi.org/10.21203/rs.3.rs-2734142/v1}.

\bibitem[Schank et~al.(2013)Schank, Berman, and Macpherson]{schank2013learning}
R.~C. Schank, T.~R. Berman, and K.~A. Macpherson.
\newblock Learning by doing.
\newblock In \emph{Instructional-design theories and models}, pages 161--181. Routledge, 2013.

\bibitem[Schmidt and Engelen(2020)]{schmidt2020}
A.~T. Schmidt and B.~Engelen.
\newblock The ethics of nudging: An overview.
\newblock \emph{Philosophy Compass}, 15\penalty0 (4):\penalty0 e12658, 2020.
\newblock \doi{https://doi.org/10.1111/phc3.12658}.
\newblock URL \url{https://compass.onlinelibrary.wiley.com/doi/abs/10.1111/phc3.12658}.

\bibitem[Sobania et~al.(2023)Sobania, Briesch, Hanna, and Petke]{sobania2023debugging}
D.~Sobania, M.~Briesch, C.~Hanna, and J.~Petke.
\newblock An analysis of the automatic bug fixing performance of chatgpt.
\newblock \emph{arXiv preprint arXiv:2301.08653}, 2023.

\bibitem[Sundin(2021)]{sundin2021nudging}
E.~Sundin.
\newblock Nudging and design friction: The impact on our decision making process.
\newblock In \emph{Conference in interaction technology and design}, page~39, 2021.

\bibitem[Sunstein(2016)]{sunstein_ethics_2016}
C.~Sunstein.
\newblock \emph{The Ethics of Influence: Government in the Age of Behavioral Science}.
\newblock Cambridge University Press, 2016.

\bibitem[Thaler and Sunstein(2009)]{thaler2009nudge}
R.~H. Thaler and C.~R. Sunstein.
\newblock \emph{Nudge: Improving decisions about health, wealth, and happiness}.
\newblock Penguin, 2009.

\bibitem[Vasconcelos et~al.(2023)Vasconcelos, J{\"o}rke, Grunde-McLaughlin, Gerstenberg, Bernstein, and Krishna]{vasconcelos2023explanations}
H.~Vasconcelos, M.~J{\"o}rke, M.~Grunde-McLaughlin, T.~Gerstenberg, M.~S. Bernstein, and R.~Krishna.
\newblock Explanations can reduce overreliance on ai systems during decision-making.
\newblock \emph{Proceedings of the ACM on Human-Computer Interaction}, 7\penalty0 (CSCW1):\penalty0 1--38, 2023.

\bibitem[Von~K{\"u}gelgen et~al.(2022)Von~K{\"u}gelgen, Karimi, Bhatt, Valera, Weller, and Sch{\"o}lkopf]{von2022fairness}
J.~Von~K{\"u}gelgen, A.-H. Karimi, U.~Bhatt, I.~Valera, A.~Weller, and B.~Sch{\"o}lkopf.
\newblock On the fairness of causal algorithmic recourse.
\newblock In \emph{Proceedings of the AAAI conference on artificial intelligence}, volume~36, pages 9584--9594, 2022.

\bibitem[Wiener and El-Yaniv(2011)]{wiener2011agnostic}
Y.~Wiener and R.~El-Yaniv.
\newblock Agnostic selective classification.
\newblock In J.~Shawe-Taylor, R.~S. Zemel, P.~L. Bartlett, F.~Pereira, and K.~Q. Weinberger, editors, \emph{Advances in Neural Information Processing Systems 24}, pages 1665--1673. Curran Associates, Inc., 2011.

\bibitem[Wilk(1999)]{Wilk1999nudges}
J.~Wilk.
\newblock \emph{Mind, Nature and the Emerging Science of Change: An Introduction to Metamorphology}, pages 71--87.
\newblock Springer Netherlands, 1999.
\newblock \doi{10.1007/978-94-017-2245-2_6}.
\newblock URL \url{https://doi.org/10.1007/978-94-017-2245-2_6}.

\bibitem[Zerilli et~al.(2022)Zerilli, Bhatt, and Weller]{zerilli2022transparency}
J.~Zerilli, U.~Bhatt, and A.~Weller.
\newblock How transparency modulates trust in artificial intelligence.
\newblock \emph{Patterns}, page 100455, 2022.

\bibitem[Zhu and Simon(1987)]{zhu1987learning}
X.~Zhu and H.~A. Simon.
\newblock Learning mathematics from examples and by doing.
\newblock \emph{Cognition and instruction}, 4\penalty0 (3):\penalty0 137--166, 1987.

\end{thebibliography}

\appendix

\section{Ethics Statement and Potential Risks}

As in the use of nudges from behavioral economics, there are critical conversations that warrant conversation around risks of selective frictions. 
While we take the stance as~\citet{thaler2009nudge} that the choice to not adjust any access to users' experiences is still a choice, there are important questions around \textit{who} is deciding when to impose frictions and on which user populations. While selective frictions could be one way to encourage critical thinking, as we begin to demonstrate here, without responsible design, they could negatively shape users' choice environments~\citep{sunstein_ethics_2016}. More pressingly, selective use of frictions may lead to disparate treatment of users, as some sub-populations of users may be frictioned on specific task instances more than others~\cite{jones2020selective}. For instance, user expertise may be distributed unequally across a protected attribute; our friction framework would then disparately friction users increasing the \textit{effort} required for some users to access the LLM output~\cite{han2020human,von2022fairness}. Systems designers ought to be aware of such disparities and take the necessary precautions when deploying frictioned access to LLMs.

% [TODO -- reference some of earlier ``risks of nudging'' work from behavioral econ.]
\begin{table*}[ht!]
\centering
\caption{\textbf{Frictioning induces minimal change in accuracy and may sway user belief in self- and model performance.} Per topic, we report average user accuracy and reported belief of expected self- and model performance (i.e., S-Belief and M-Belief respectively). Error bars indicate standard error over participants.}
\resizebox{\textwidth}{!}{%

\begin{tabular}{l|ccc|ccc|ccc|ccc}
\toprule
 & \multicolumn{3}{c|}{US Foreign Policy} & \multicolumn{3}{c|}{Math} & \multicolumn{3}{c|}{CS} & \multicolumn{3}{c}{Bio} \\

Variant & Acc & S-Belief & M-Belief & Acc & S-Belief & M-Belief & Acc & S-Belief & M-Belief & Acc & S-Belief & M-Belief \\
\midrule\hline
Baseline & 0.40 \footnotesize{$\pm$ 0.04} & 0.35 \footnotesize{$\pm$ 0.07} & 0.53 \footnotesize{$\pm$ 0.06} & 0.67 \footnotesize{$\pm$ 0.06} & 0.47 \footnotesize{$\pm$ 0.08} & 0.63 \footnotesize{$\pm$ 0.08} & 0.55 \footnotesize{$\pm$ 0.04} & 0.25 \footnotesize{$\pm$ 0.06} & 0.58 \footnotesize{$\pm$ 0.07} & 0.67 \footnotesize{$\pm$ 0.05} & 0.40 \footnotesize{$\pm$ 0.07} & 0.60 \footnotesize{$\pm$ 0.06} \\
Friction & 0.46 \footnotesize{$\pm$ 0.05} & 0.47 \footnotesize{$\pm$ 0.06} & 0.45 \footnotesize{$\pm$ 0.06} & 0.66 \footnotesize{$\pm$ 0.05} & 0.58 \footnotesize{$\pm$ 0.07} & 0.57 \footnotesize{$\pm$ 0.08} & 0.58 \footnotesize{$\pm$ 0.04} & 0.36 \footnotesize{$\pm$ 0.06} & 0.53 \footnotesize{$\pm$ 0.07} & 0.67 \footnotesize{$\pm$ 0.04} & 0.48 \footnotesize{$\pm$ 0.06} & 0.55 \footnotesize{$\pm$ 0.06} \\
\bottomrule
\end{tabular}

}

\label{agg_compare}
\end{table*}

\begin{figure}[t]
    \centering
    \includegraphics[width=\linewidth]{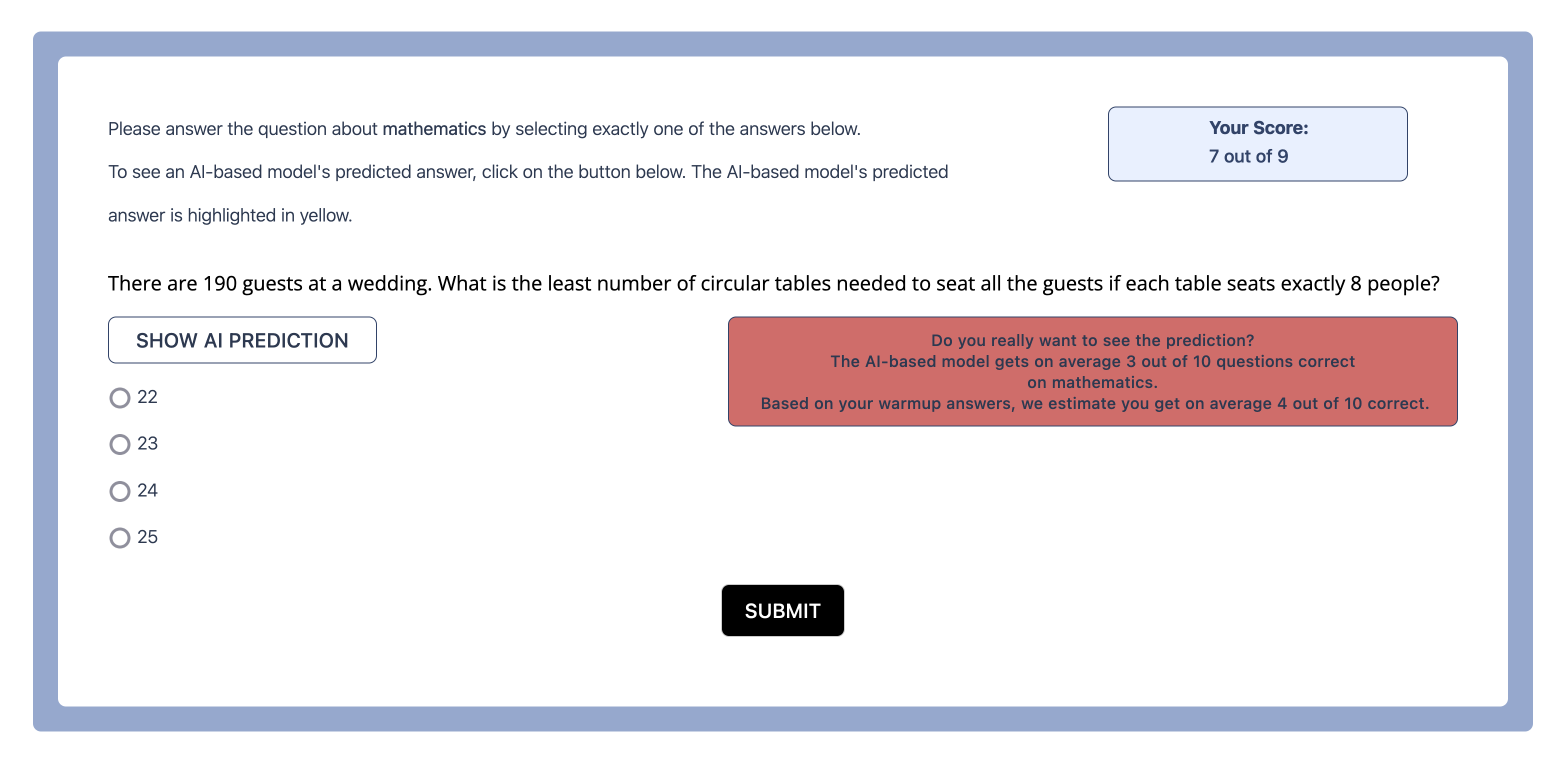}
    \caption{Interface for selective friction; here the user scored higher than the model in their pre-quiz on mathematics. If the user presses the ``Show AI Prediction'' button (first button) they are then presented with the red block (second button) which they are forced to click if they still want to see the prediction.}
    \label{fig:friction-interface}
\end{figure}

\section{Related Work}

In this work, we focus on LLM-assisted user interactions and decision-making~\citep{green2019principles, lai2021towards}.
Prior studies studying these contexts have shown the tendency for humans to overrely on AI support~\citep{joshi2023machine, chen2023understanding, vasconcelos2023explanations}.
As such, recent works have considered adapting when AI support is provided to users:
\citet{ma2023should} fit a decision tree to offline users' decisions to decide when to show AI support to users, \citet{buccinca2024towards} use offline reinforcement learning to estimate if AI support would be helpful, and \citet{bhatt2023learning} employ online learning techniques to personalize a decision support policy to individuals.
Relatedly, others have considered selectively delegating entire tasks to the AI model~\citep{madras2018predict,mozannar2020consistent,lai2022human, wiener2011agnostic,geifman2017selective}.
In many cases, it may not be possible (or desirable even if possible) to have an LLM make the final decision, and in others, it may not be justifiable to withhold user access to LLMs
These constraints motivate our study of frictions, which permit continued model access but require more effort on the user's end to procure access.

Our work builds on the wealth of prior research into the design and effect of nudges on human behavior~\citep{thaler2009nudge, schmidt2020, hummel2019}. The notion of nudges is increasingly permeating machine learning, whether in the use of techniques from machine learning to design nudges~\citep{callaway2023optimal} or nudging to support more appropriate use of AI systems~\citep{li2024decoding, buccinca2021}.
Relatedly,  ``microboundaries''~\citep{cox2016microboundry}, are small, intentional barriers integrated into user interfaces to promote more mindful interactions. Microboundaries can reduce the likelihood of users making errors or engaging in habitual, potentially harmful behaviors by interrupting their flow and requiring them to take an additional step before proceeding. There are several potential advantages of microboundaries (a la warning signals) as a type of design friction~\citep{sundin2021nudging,mejtoft2019design,alon2023human}.
To our knowledge, we are the first work to explore selective frictioning of LLM use.

\section{Additional Details on Human Experiments}
\label{sec:human-exp-details}

\subsection{Participant Recruitment} We provide additional details on our user study. Participants receive the quiz for all conditions. The same questions are presented across both conditions, selected from three batches of $60$ questions, as in~\citep{bhatt2023learning}. The ``test'' phase involves $10$ questions per topic. Participants are provided feedback as to whether they (and the model, if seen) are correct after each test trial. Feedback is not given in the quiz phase. Participants are paid at a base rate of $\$9$ per hour for an expected $30$ minute experiment with an optional bonus up to $\$10$ per hour for correct answers; we apply the bonus to all participants. All data is anonymized, and participants provided informed consent before beginning the study.

\subsection{Eliciting Perceived Self- and Model-Ability}
\label{self-model-ability}

At the end of the study, users are presented with a questionarre asking them to judge their own and the model's ability per topic. Specifically, we asked, for each topic: ``Out of 100 questions on TOPIC, how many do you think \textit{the AI} would get correct?'' and ``Out of 100 questions on TOPIC, how many do you think \textit{you} could get correct (without the help of the AI-based model)?'' For each question, users responded on an slider ranging from $0$ to $100$.

\subsection{LLM Predictions} 
\label{llm-pred-details}

We use the same model predictions as in ~\citep{bhatt2023learning}, which were sampled from InstructGPT3.5 \texttt{text-davinci-003}~\citep{ouyang2022training}; however, we randomly dampen model performance for the foreign policy and computer science topics such that the models achieve 30\% and 60\% performance on each, respectively. The model achieves approximately 30\% and 90\% performance on mathematics and biology. MMLU is challenging for humans~\citep{hendryckstest2021, bhatt2023learning, mozannar2024effective}; our selective friction is triggered by user performance relative to the model's average performance on a topic -- if the latter is too high, the friction will not be triggered. Accordingly, we expect with the dampening that we should have high trigger rates for mathematics and foreign policy, moderate rates for computer science, and low rates for biology -- enabling us to study user click behavior across a range of model performances and settings wherein in some cases it is indeed rationale to rely on the model whereas in others, it may be disastrous for a user to regularly rely on the model (e.g., in elementary mathematics). 

\subsection{User Study Interface} 
\begin{figure*}
    \centering
    \includegraphics[width=0.7\textwidth]{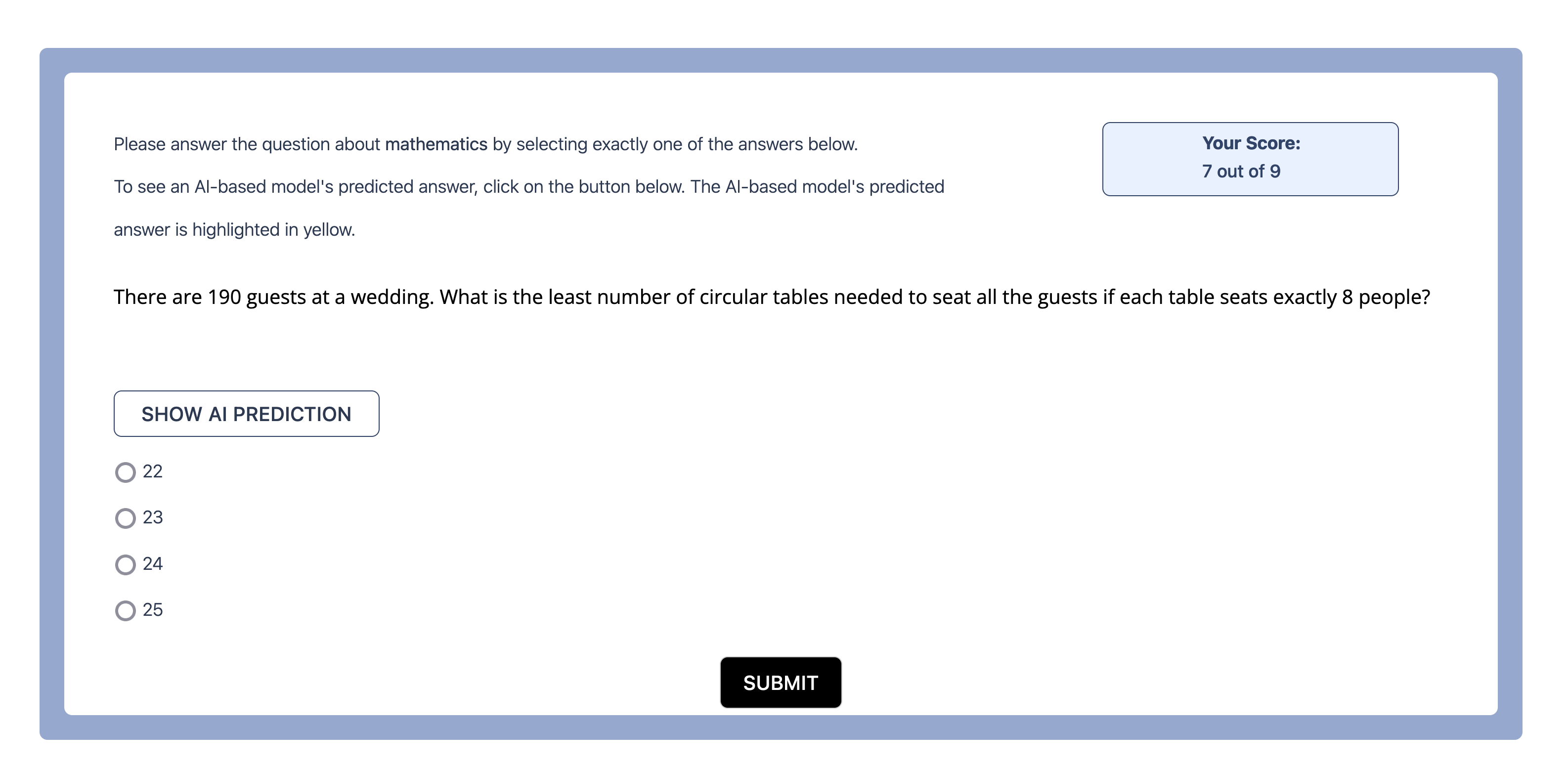}
    \caption{Example interface that the user is presented with for each MMLU question, where they have the option to click a button to query the AI.}
    \label{fig:btn-click-interface}
\end{figure*}

\begin{figure*}
    \centering
    \includegraphics[width=0.7\textwidth]{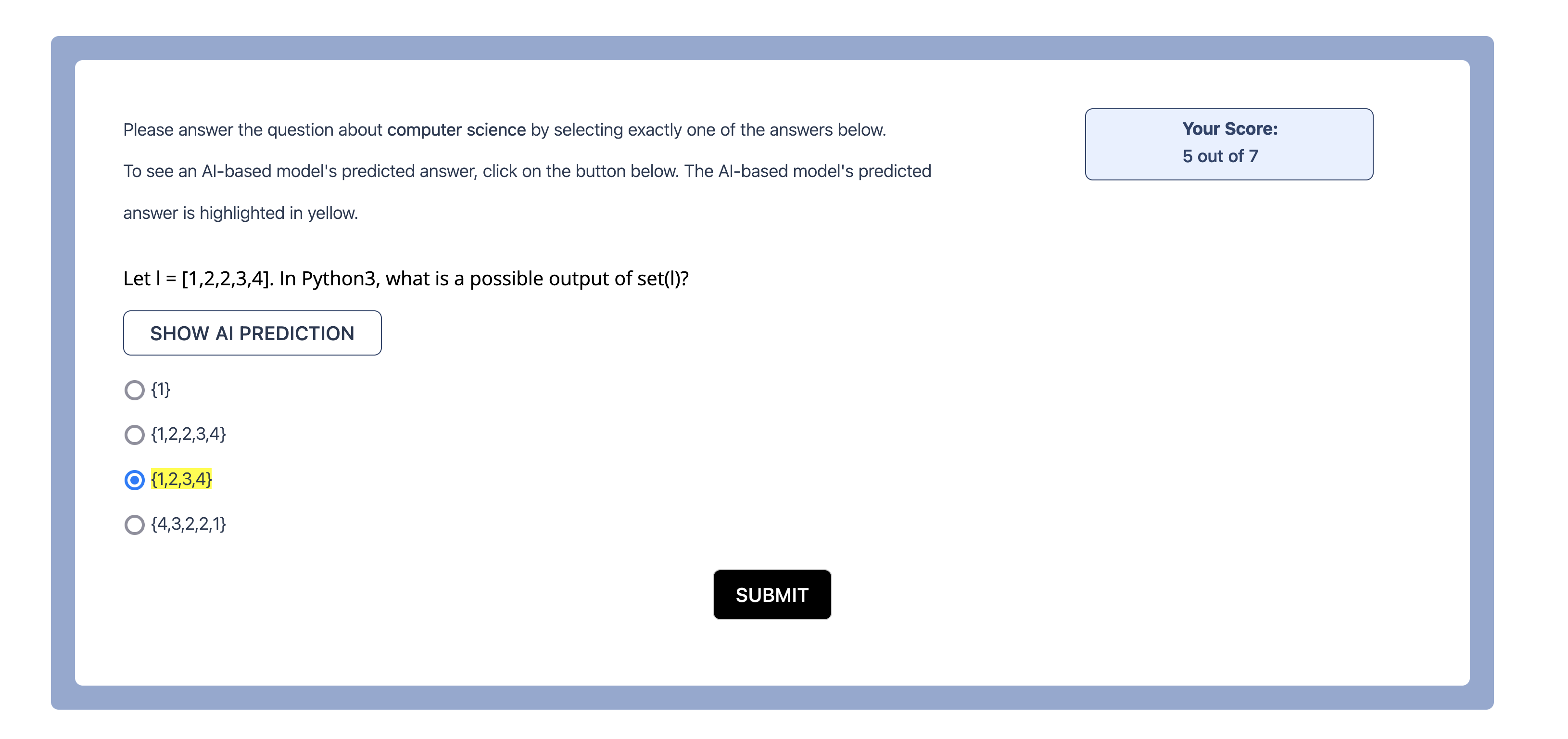}
    \caption{Example interface after the user has clicked the button to query the AI and the model prediction is shown via highlight.}
    \label{fig:btn-click-interface-llm-shown}
\end{figure*}

We include example screenshots of the button interface and friction in Figures \ref{fig:btn-click-interface} and \ref{fig:friction-interface}, respectively. When the user does click through to the LLM prediction, it is displayed as in \ref{fig:btn-click-interface-llm-shown}, following~\citet{bhatt2023learning}.

\section{Additional Human Experiment Results}
\label{sec:additional-results}

We include additional exploratory investigations into user behavior. We report average time spent for each topic in Table \ref{time_spent}. We observe that the average time spent per problem appears to increase across topics.%, the difference is only statistically significant for foreign policy.

\begin{figure*}
    \centering
    \includegraphics[width=0.7\textwidth]{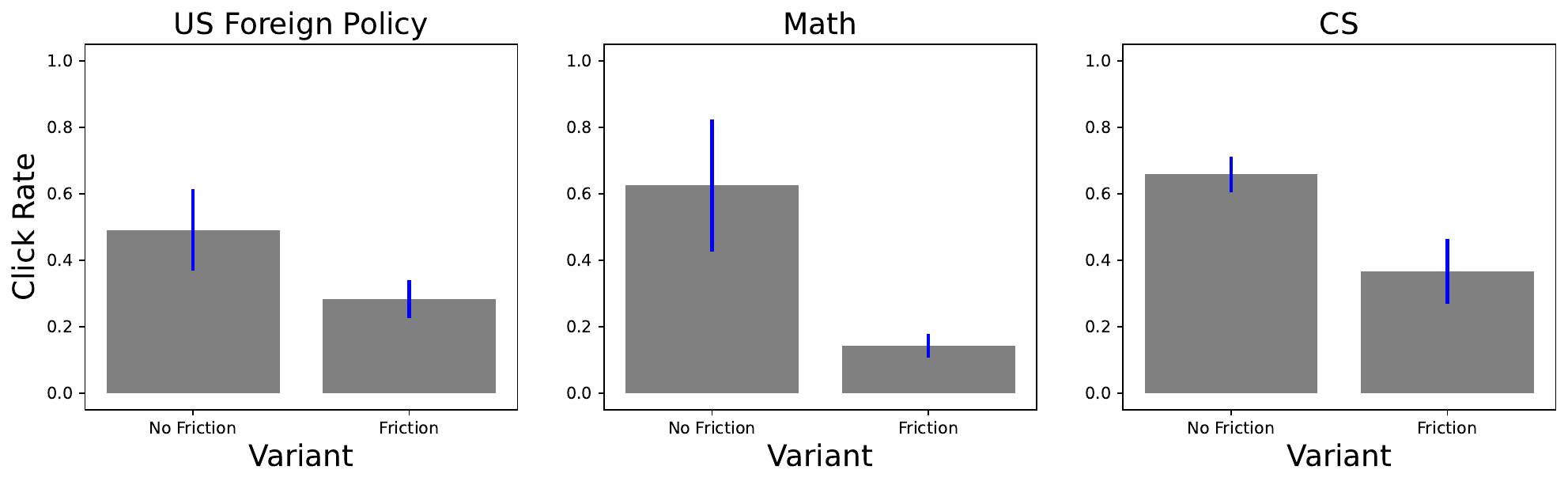}
    \caption{Comparing per-topic click rates people within the friction condition ($N=53$) who did or did not see a friction. No one was frictioned for biology.} %Top: users did not receive friction for that topic; bottom: users did see the friction for a topic.}
    \label{fig:num-click-decomp}
\end{figure*}

We also decompose user behavior in Figure~\ref{fig:num-click-decomp} within the frictioned condition according to whether the participant received a friction on that topic or not. Here, we can more clearly see that click rates decrease for participants who are explicitly frictioned. However, we caveat these results in that the friction intervention itself induces biases in the user samples across the groups. Participants only see a friction if they are necessarily better than the model; hence, users may already be inclined to click less often. % hence, we would expect the participants to have higher base accuracy across the topics. 

\begin{table*}[ht!]
\centering
\caption{Average time (seconds) per user per topic per question. We do not generally observe a significant difference in the amount of time spent as a result of the selective frictioning.}
\begin{tabular}{l|c|c|c|c}
\toprule
 & {US Foreign Policy} & {Math} & {CS} & {Bio} \\

Variant & Time (sec) & Time (sec) & Time (sec) & Time (sec) \\
\midrule\hline
Baseline & 18.53 $\pm$ 1.59 & 23.23 $\pm$ 3.31 & 18.58 $\pm$ 2.50 & 22.80 $\pm$ 8.49 \\
Friction
Condition & 23.29 $\pm$ 2.49 & 27.25 $\pm$ 3.15 & 18.85 $\pm$ 1.74 & 23.48 $\pm$ 3.72 \\
\bottomrule
\end{tabular}

\label{time_spent}
\end{table*}

% \begin{table*}[ht!]
% \centering
% \resizebox{\textwidth}{!}{%
% \begin{tabular}{l|cc|cc|cc|cc|}
% \toprule
%  & \multicolumn{2}{c|}{US Foreign Policy} & \multicolumn{2}{c|}{Math} & \multicolumn{2}{c|}{CS} & \multicolumn{2}{c|}{Bio} \\
% Variant & Acc & Click Rate & Acc & Click Rate & Acc & Click Rate & Acc & Click Rate \\
% \midrule
% \hline
% No Friction & 0.37 $\pm$ 0.08 & 0.49 $\pm$ 0.22 & 0.50 $\pm$ 0.21 & 0.62 $\pm$ 0.33 & 0.57 $\pm$ 0.05 & 0.66 $\pm$ 0.09 & 0.67 $\pm$ 0.04 & 0.48 $\pm$ 0.09 \\Friction & 0.48 $\pm$ 0.05 & 0.28 $\pm$ 0.10 & 0.67 $\pm$ 0.05 & 0.14 $\pm$ 0.06 & 0.58 $\pm$ 0.08 & 0.37 $\pm$ 0.17 & - & - \\
% \bottomrule
% \end{tabular}
% }
% \caption{Comparing people within the friction condition ($N=53$) who did or did not see a friction. Top: users did not receive friction for that topic; bottom: users did see the friction for a topic.}
% \label{decomp_win_fric_compare}
% \end{table*}

\subsection{Example User Responses}
\label{post-resps}

We asked participants in a post-survey questionarre what factors led them to click. We include a few exemplary responses below: 

\begin{itemize}
    % asked about why click, for the baseline conditions 
    \item \texttt{``On questions where I didn't immediately know the answer, I clicked to see the model predicted answer in case it would help me. On questions where I didn't know the answer at all, I clicked to see the model predicted answer to help me know where to start.''}
    \item \texttt{``If I simply had no idea what the answer was to a question, I would click to see what the prediction would say and then decide if I wanted to go along with it or not. The other instance would be if I was stuck between two choices, I'd click the button to see if the prediction was the same answer as mine or not to have some sort of confirmation.''}
    \item \texttt{``not knowing anything about the topic or having any idea as to the right answer i clicked on the the ai button hoping it would help or know more than me. I also clicked it a couple times when i thought i knew the answer but wasnt 100\% sure and if it chose the same as me i felt more confident''}
    \item \texttt{``Unsure if I had the right answer and not having enough knowledge or not having used knowledge of the subject for up to 30 years''}

    % from frictioned participants

    \item \texttt{``I used the prediction button if I felt unsure of the answer or if I wanted to feel more assured of my own answer.''}

    \item \texttt{``At first it was to help me with answering the question, then I realized the AI gave wrong answers as well, so for the ones I was sure of the answer I still clicked to see what it would show.It was very satisfactory to see I got it right when AI got it wrong, but very disappointing when AI gave me wrong answer when I didn't know the correct answer''}

    \item \texttt{``If I was unsure of the correct answer (or second guessing myself), I checked the model prediction to see if the AI model aligned with what I was thinking''}
\end{itemize}

When asked why they clicked, we noted that a few participants did report that they were simply curious: 

\begin{itemize}
    % from the baseline
    \item \texttt{``In instances where I doubted my answer, I clicked to see the A.I model prediction. After getting an answer wrong I had an urge to click on the prediction. When I had no clue what to answer, I clicked the button. Sometimes, especially, for the math questions I clicked out of curiosity since I observed the A.I often got the answers wrong.''}
    \item \texttt{``I was mainly interested to see what it thought the answer was, independent if I thought I knew the answer or not. I had to look.''}

    % from frictioned participant 

    \item \texttt{``I am not familiar with the terminology and am curious how AI responds.''}
\end{itemize}

We also asked participants why they chose \textit{not} to click:

\begin{itemize}

% baseline condition 
\item \texttt{``I already 100\% knew the answer so I didn't wait to see what the model predicted answer was.''}
\item \texttt{``I thought it would struggle to answer some of the more linguistically complicated questions correctly (like the ones that asked which of these is not true and 3 things are and 1 is not). I also thought for the most basic math problems (like things that were essentially a single computation), there wasn't really a need.''}
\item \texttt{``I didn't click the button for answers where I was highly confident. (I suppose it probably wouldn't have hurt to click it, but it would take a little pride out of it if the AI was correct too...)''}

% from friction condition 
    \item \texttt{``I like to challenge myself naturally, I'd prefer to make a good guess and be wrong to learn from it than to just look up the answer (or in this case, use AI) and immediately forget. I'd argue this is a case of disconnect between an internal effort and reward system.''}
    \item \texttt{``If I was fairly certain I already knew the answer, then I did not click it.''}
    \item \texttt{``The prompt [the friction] that came up was a major factor. It provided me with data that I was twice as likely to get it right than if I had not used it at all.''}
    \item \texttt{``If I had a good idea or felt that I could answer quickly, I disregarded the AI suggestion since that would have wasted time.''}
    \item \texttt{``[in the friction condition] Being told the AI would mostly get it wrong and confidence in my own ability to answer''}
    \item \texttt{``[in the friction condition] Being told the AI would mostly get it wrong and confidence in my own ability to answer''}
    \item \texttt{``I felt confident in my answer, and knowing that I was a better predictor in certain categories than the model''}
    \item \texttt{``If I thought I knew the answer to a question, I felt no reason to consult the AI. However, I did check the prediction when I had an answer, but I was unsure of it.''}
    
\end{itemize}

% \begin{figure*}
%     \centering
%     \includegraphics[width=0.7\textwidth]{figures/friction_math.png}
%     \caption{Sample interface of selective friction; here the user scored higher than the model in their pre-quiz on mathematics. Hence, they are forced to click a second button if they still want to see the prediction.}
%     \label{fig:friction-interface}
% \end{figure*}

\section{Experiment Instructions}

\begin{figure*}
    \centering
    \begin{mdframed}[leftmargin=10pt,rightmargin=10pt]
    \includegraphics[width=0.85\linewidth]{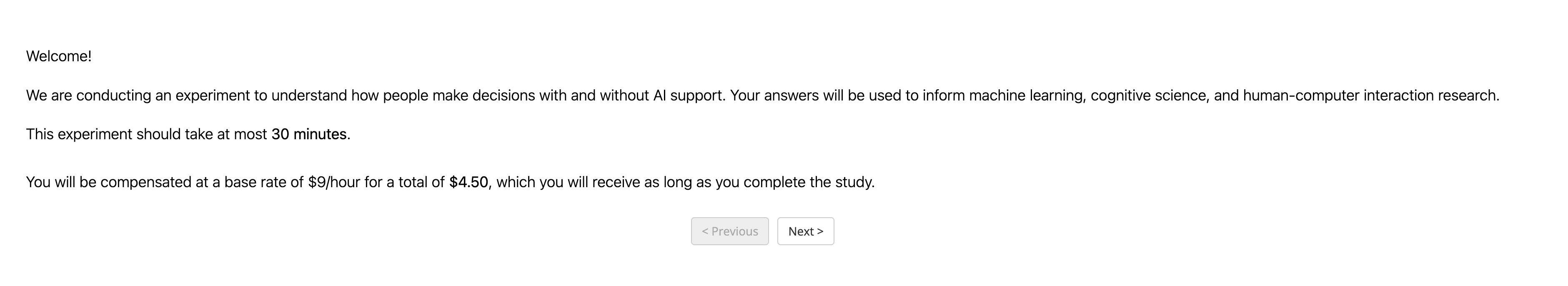}
    \end{mdframed}
    \vspace{0.5mm}
    \begin{mdframed}[leftmargin=10pt,rightmargin=10pt]
    \includegraphics[width=0.85\linewidth]{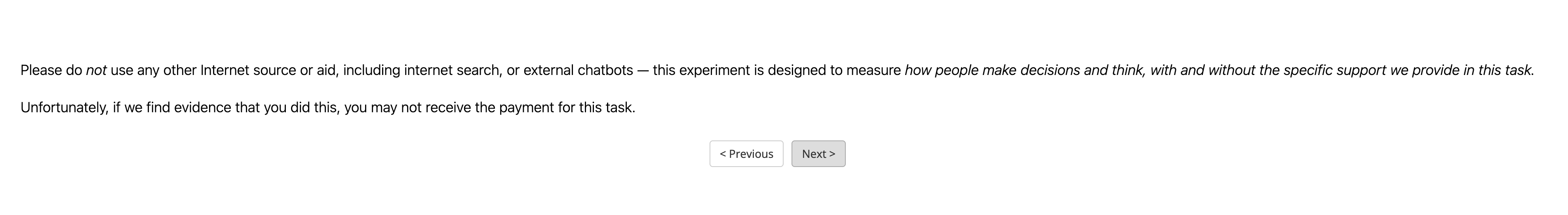}
    \end{mdframed}
    \vspace{0.5mm}
    \begin{mdframed}[leftmargin=10pt,rightmargin=10pt]
    \includegraphics[width=0.85\linewidth]{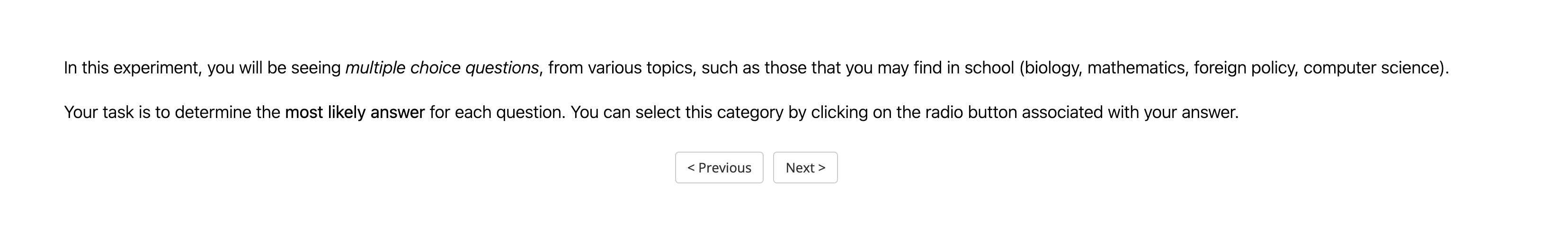}
    \end{mdframed}
    \vspace{0.5mm}
    \begin{mdframed}[leftmargin=10pt,rightmargin=10pt]
    \includegraphics[width=0.85\linewidth]{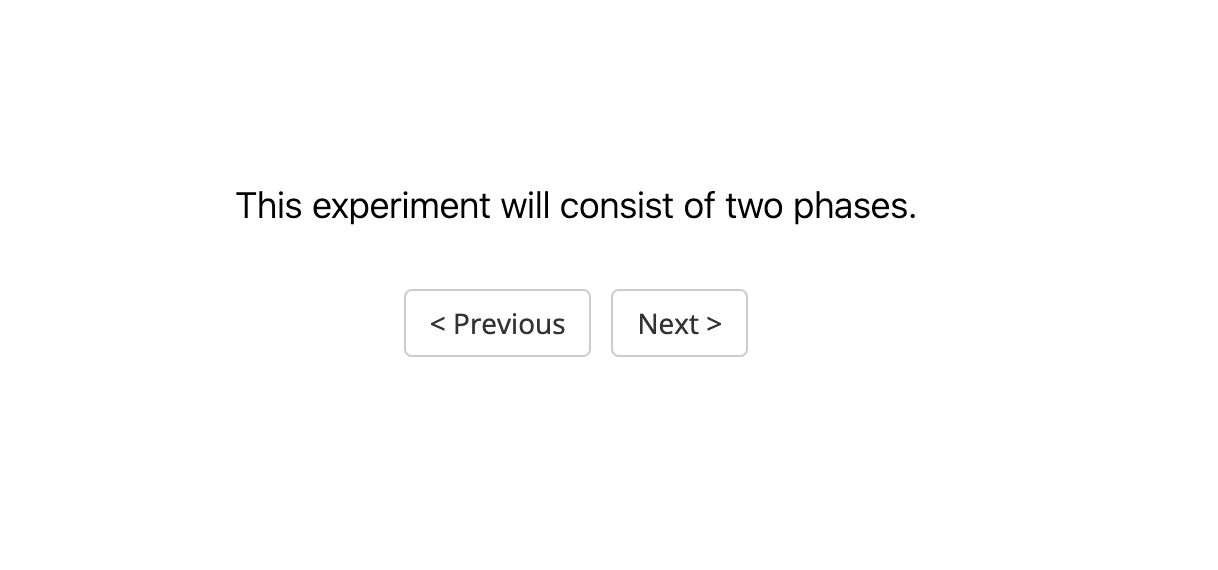}
    \end{mdframed}
    \vspace{-2mm}
    \caption{Experiment instructions.}
    \label{fig:experiment-instructions}
\end{figure*}

\begin{figure*}
    \centering
    \begin{mdframed}[leftmargin=10pt,rightmargin=10pt]
    \includegraphics[width=0.65\linewidth]{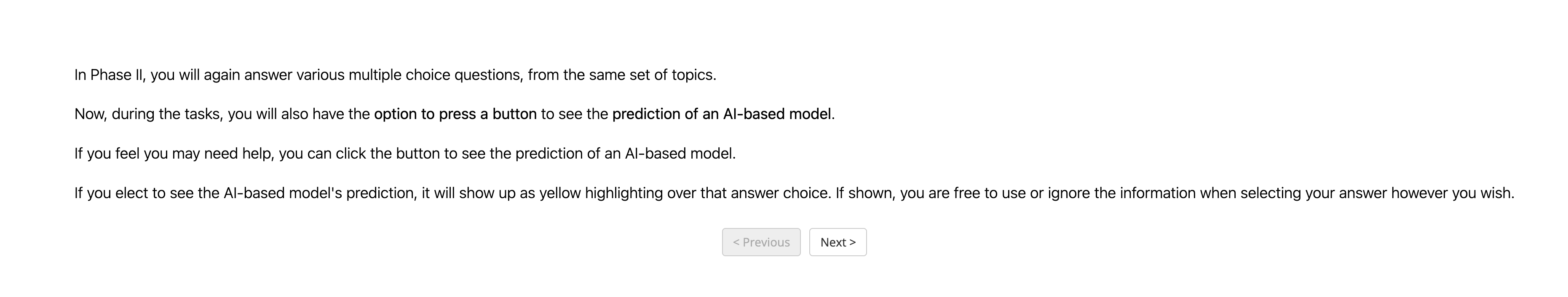}
    \end{mdframed}
    \vspace{0.5mm}
    \begin{mdframed}[leftmargin=10pt,rightmargin=10pt]
    \includegraphics[width=0.65\linewidth]{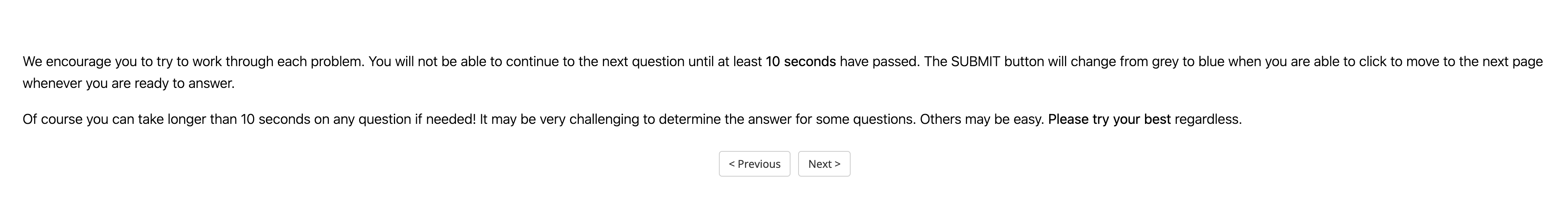}
    \end{mdframed}
    \vspace{0.5mm}
    \begin{mdframed}[leftmargin=10pt,rightmargin=10pt]
    \includegraphics[width=0.7\linewidth]{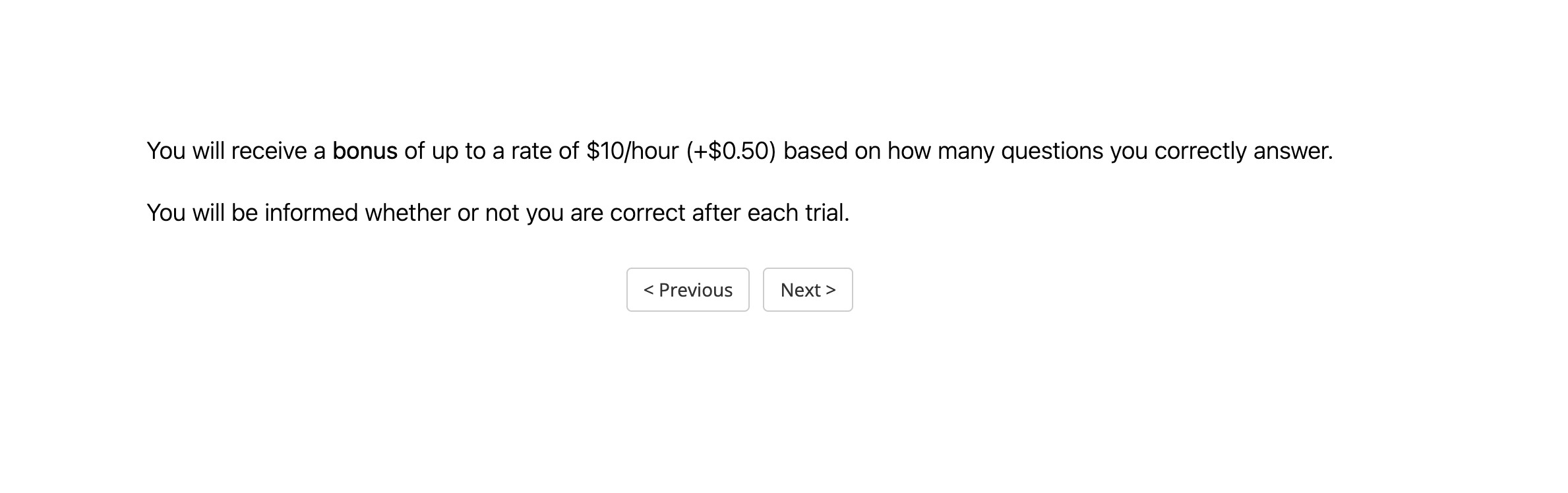}
    \end{mdframed}
    \vspace{0.5mm}
    \begin{mdframed}[leftmargin=10pt,rightmargin=10pt]
    \includegraphics[width=0.65\linewidth]{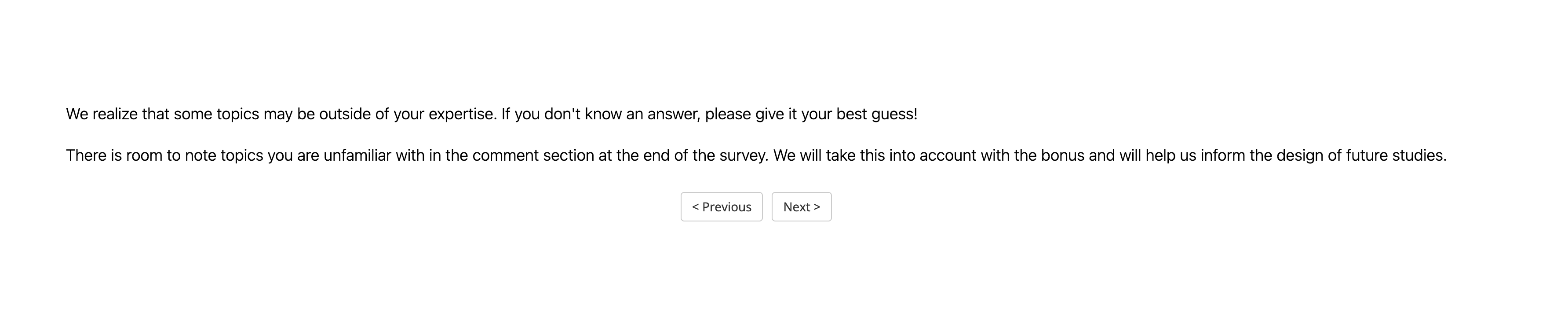}
    \end{mdframed}
    \begin{mdframed}[leftmargin=10pt,rightmargin=10pt]
    \includegraphics[width=0.65\linewidth]{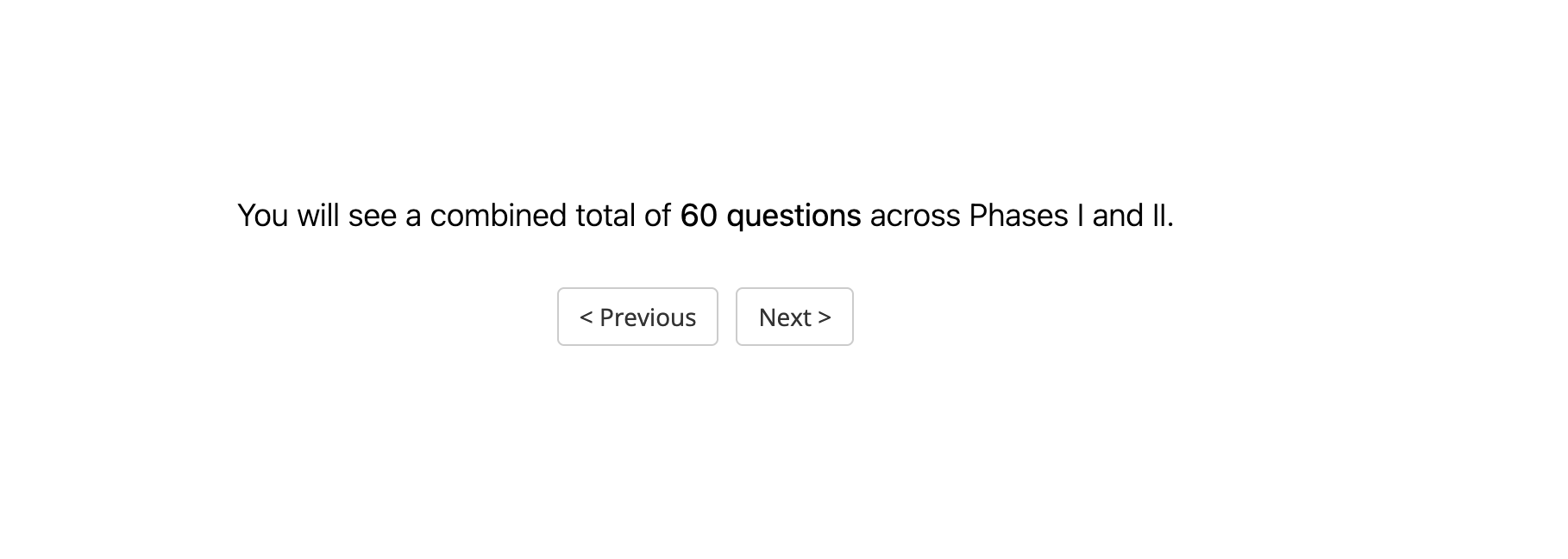}
    \end{mdframed}
    \begin{mdframed}[leftmargin=10pt,rightmargin=10pt]
    \includegraphics[width=0.65\linewidth]{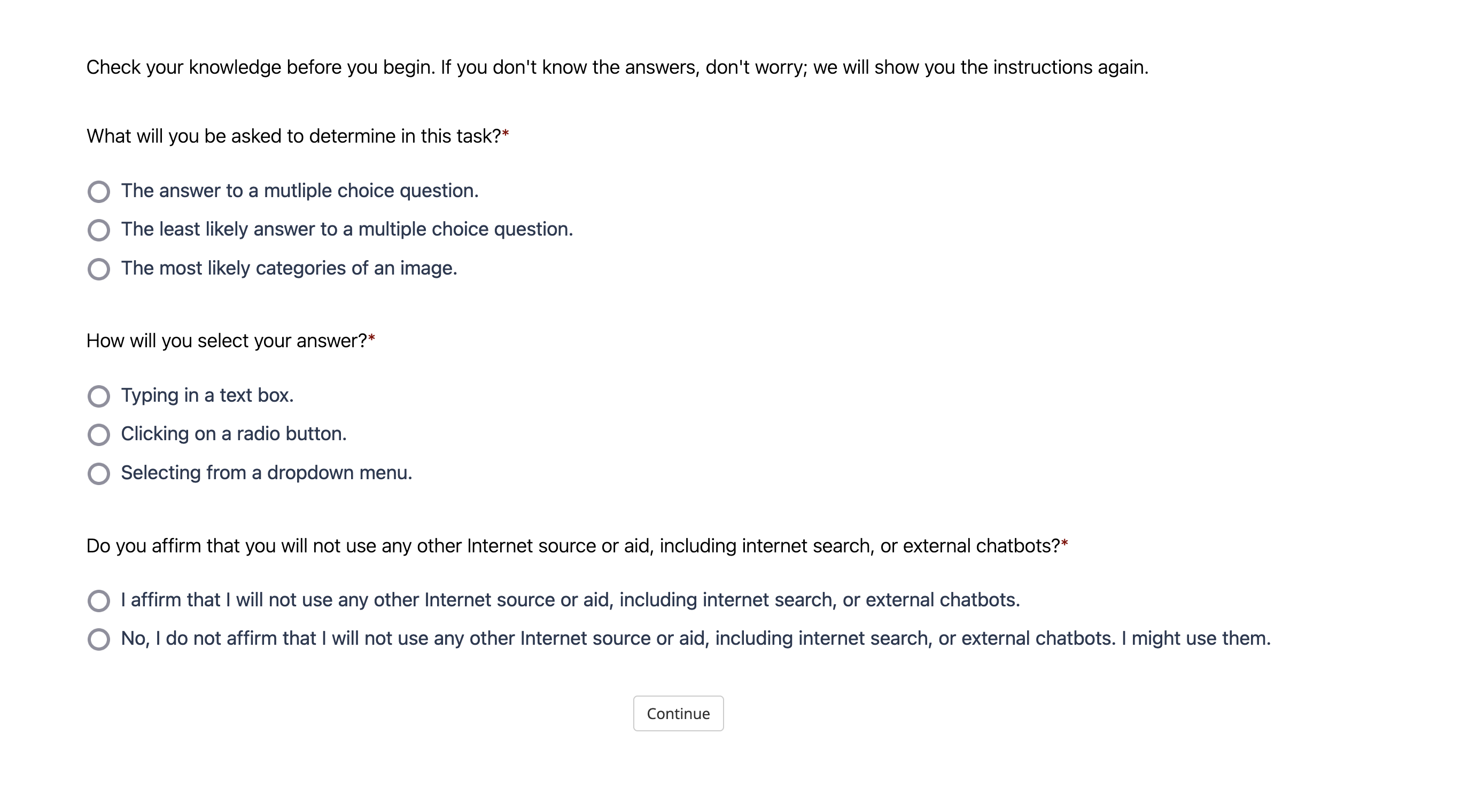}
    \end{mdframed}
    \begin{mdframed}[leftmargin=10pt,rightmargin=10pt]
    \includegraphics[width=0.65\linewidth]{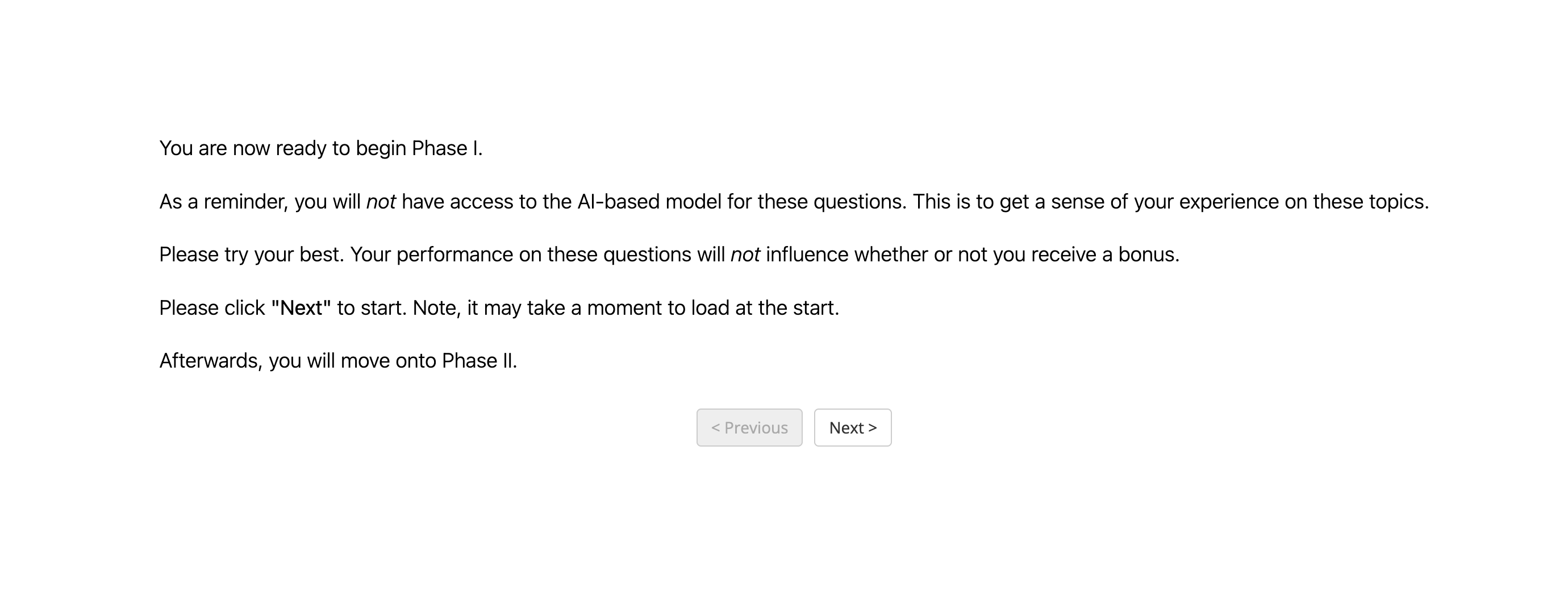}
    \end{mdframed}
    \caption{Experiment instructions (continued).}
    \label{fig:experiment-instructions-2}
\end{figure*}

\begin{figure*}
    \centering
    \begin{mdframed}[leftmargin=10pt,rightmargin=10pt]
    \includegraphics[width=0.65\linewidth]{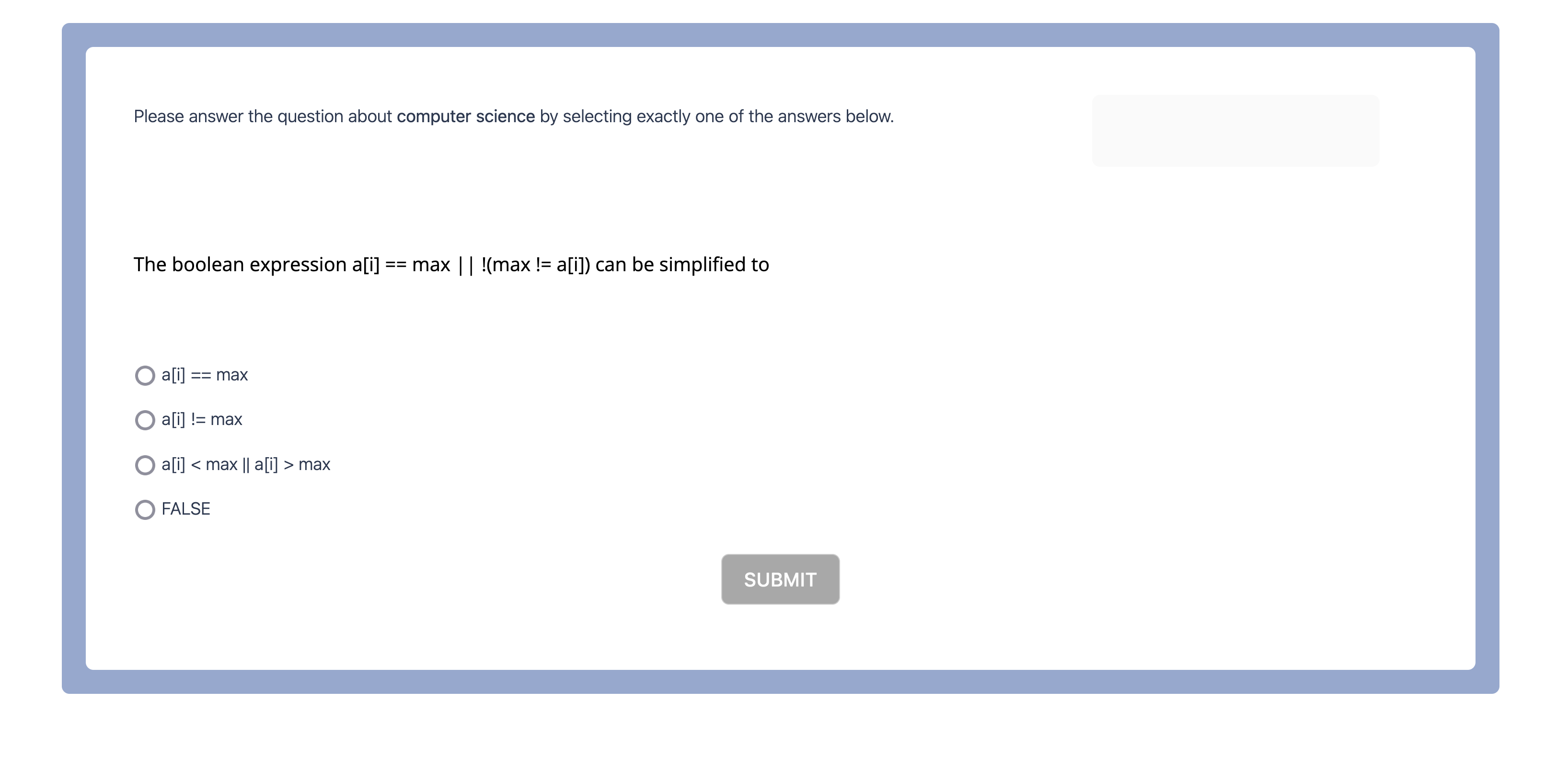}
    \end{mdframed}
    \begin{mdframed}[leftmargin=10pt,rightmargin=10pt]
    \includegraphics[width=0.65\linewidth]{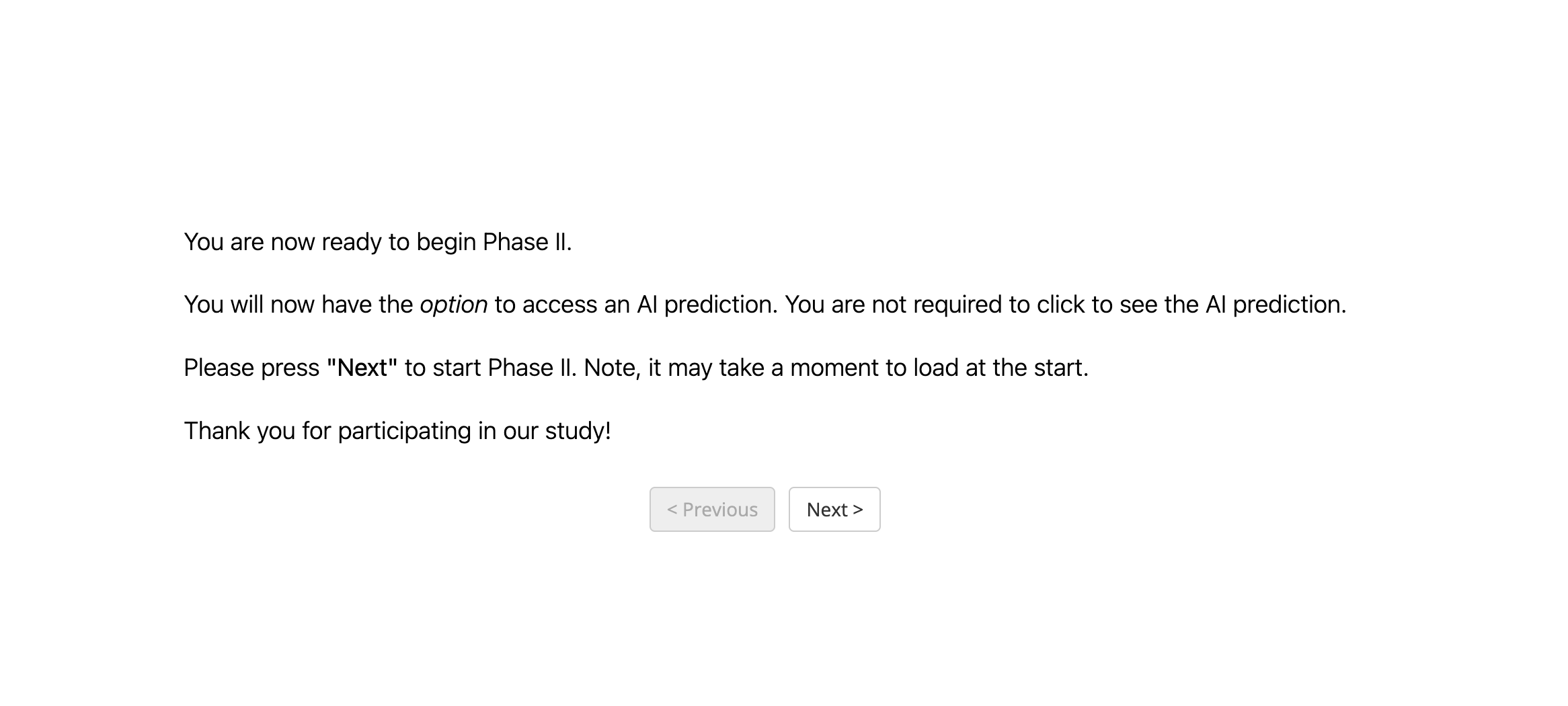}
    \end{mdframed}
    \begin{mdframed}[leftmargin=10pt,rightmargin=10pt]
    \includegraphics[width=0.65\linewidth]{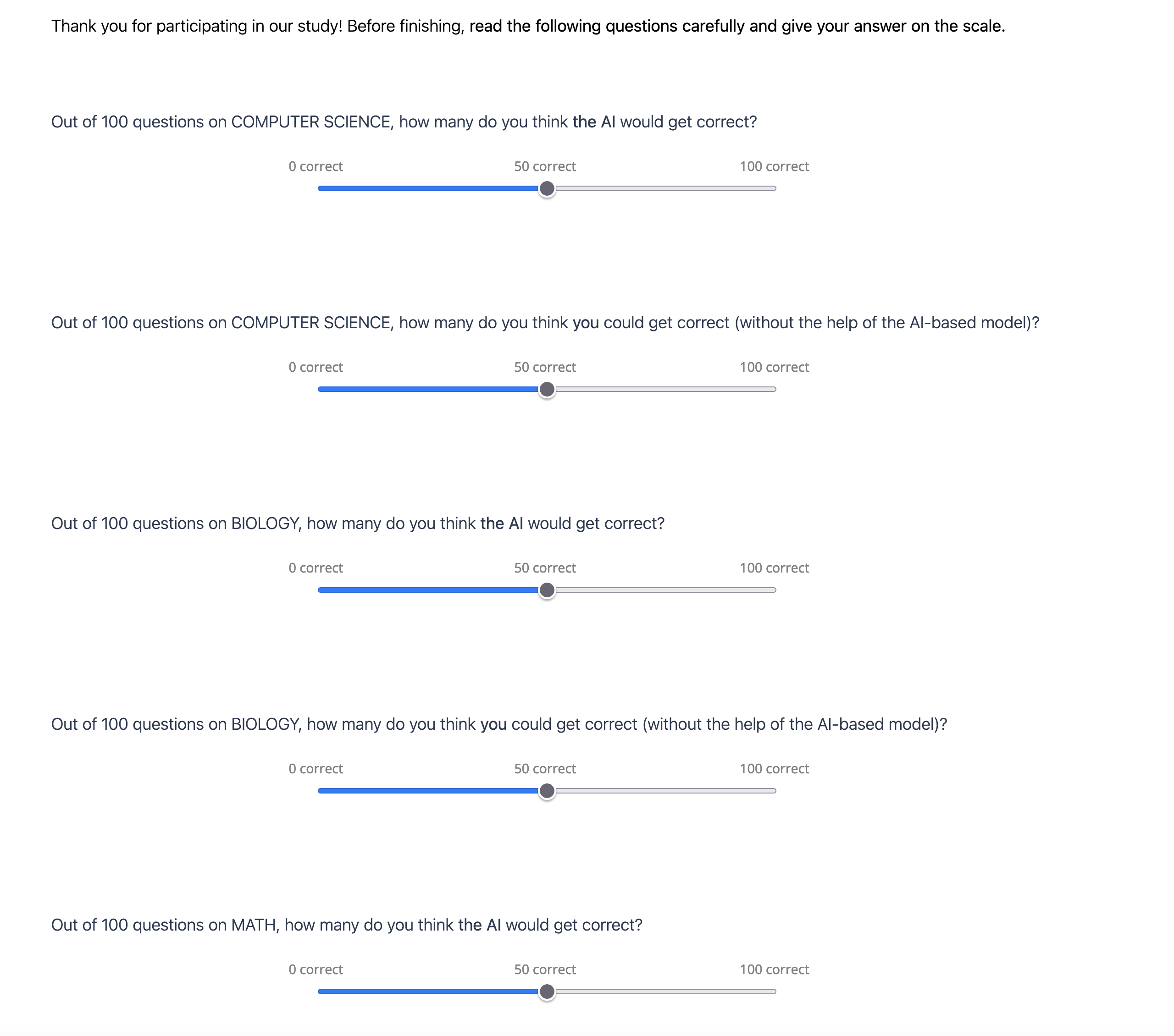}
    \end{mdframed}
    \begin{mdframed}[leftmargin=10pt,rightmargin=10pt]
    \includegraphics[width=0.65\linewidth]{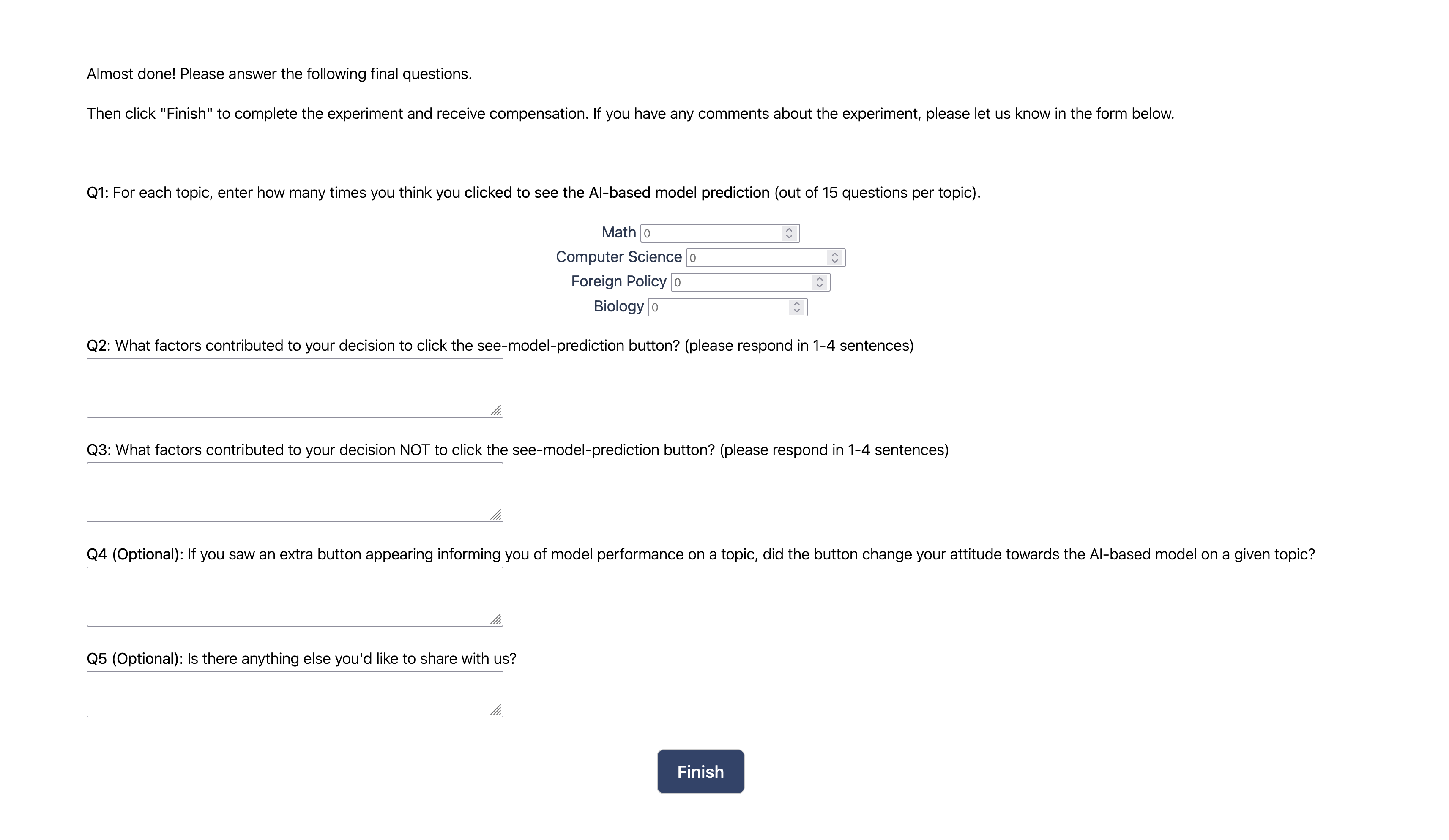}
    \end{mdframed}
    \caption{Experiment instructions (continued). The screen of Phase II is the official trial as presented in Figures \ref{fig:btn-click-interface}, \ref{fig:btn-click-interface-llm-shown}, and  \ref{fig:friction-interface}, respectively. The Phase I interface follows the same format, but no model access is permitted.}
    \label{fig:experiment-instructions-3}
\end{figure*}

We include instructions presented in our user study in Figures \ref{fig:experiment-instructions}, \ref{fig:experiment-instructions-2}, and \ref{fig:experiment-instructions-3}. 

\end{document}